\documentclass[sigconf,nonacm]{acmart}
\AtBeginDocument{%
  }

\setcopyright{acmlicensed}
\copyrightyear{2018}
\acmYear{2018}
\acmDOI{XXXXXXX.XXXXXXX}
\acmConference[Conference acronym 'XX]{Make sure to enter the correct
  conference title from your rights confirmation email}{June 03--05,
  2018}{Woodstock, NY}
\acmISBN{978-1-4503-XXXX-X/2018/06}

\usepackage{pgfplots}
\usepackage{pgfplotstable} %
\pgfplotsset{compat=1.18}
\usepackage{subcaption}
\usetikzlibrary{fpu}
\usetikzlibrary{pgfplots.colorbrewer}
\usetikzlibrary{calc}
\usepackage{multirow}
\usepgfplotslibrary{groupplots}
\usepackage[T1]{fontenc}
\begin{document}

\title{Scaling Laws for Embedding Dimension in Information Retrieval}

\author{Julian Killingback$^*$, Mahta Rafiee$^*$, Madine Manas, and Hamed Zamani}

\affiliation{%
  \institution{Center for Intelligent Information Retrieval}
  \institution{University of Massachusetts Amherst}
  \city{Amherst}
  \state{Massachusetts}
  \country{USA}
}
\email{{jkillingback, mrafiee, mmadine, zamani}@cs.umass.edu}

\renewcommand\thefootnote{\fnsymbol{footnote}}

\renewcommand{\shortauthors}{Killingback et al.}

\begin{abstract}
Dense retrieval, which encodes queries and documents into a single dense vector, has become the dominant neural retrieval approach due to its simplicity and compatibility with fast approximate nearest neighbor algorithms. As the tasks dense retrieval performs grow in complexity, the fundamental limitations of the underlying data structure and similarity metric---namely vectors and inner-products---become more apparent. Prior recent work has shown theoretical limitations inherent to single vectors and inner-products that are generally tied to the embedding dimension. Given the importance of embedding dimension for retrieval capacity, understanding how dense retrieval performance changes as embedding dimension is scaled is fundamental to building next generation retrieval models that balance effectiveness and efficiency. In this work, we conduct a comprehensive analysis of the relationship between embedding dimension and retrieval performance. Our experiments include two model families and a range of model sizes from each to construct a detailed picture of embedding scaling behavior. We find that the scaling behavior fits a power law,  allowing us to derive scaling laws for performance given only embedding dimension, as well as a joint law accounting for embedding dimension and model size. Our analysis shows that for evaluation tasks aligned with the training task, performance continues to improve as embedding size increases, though with diminishing returns. For evaluation data that is less aligned with the training task, we find that performance is less predictable, with performance degrading with larger embedding dimensions for certain tasks. We hope our work provides additional insight into the limitations of embeddings and their behavior as well as offers a practical guide for selecting model and embedding dimension to achieve optimal performance with reduced storage and compute costs.
\end{abstract}


\maketitle

\footnotetext{$^*$ Both authors contributed equally to this research.}

\begin{figure*}[ht]     
    
    \centering
    \definecolor{DataPointColor}{HTML}{10402d} %
    \definecolor{FitLineColor}{HTML}{2aa876}
    \definecolor{SoftGridColor}{HTML}{EEEEEE}  %
    \definecolor{AxisFrameColor}{HTML}{424242} %

    \begin{subfigure}[b]{0.245\textwidth}
        \centering
        \begin{tikzpicture}
            \newcommand{\EqA}{9.76706911832506}
            \newcommand{\EqAlpha}{1.7600137705003502}
            \newcommand{\EqE}{0.023524591650654365}
                        
            \begin{axis}[
                title={\small BERT - MSMARCO Dev},
                ylabel={\small Contrastive Entropy},
                xmode=log, log basis x={2},
                grid=major,
                grid style={solid, SoftGridColor},
                width=0.75\textwidth, %
                height=2.0cm,
                scale only axis,
                tick pos=left, 
                tick label style={font=\footnotesize},
                title style={font=\footnotesize\scshape, yshift=-0.75ex},
                y tick label style={/pgf/number format/fixed, font=\footnotesize, /pgf/number format/assume math mode=true},
                scaled y ticks=false,
                xtick={32, 128, 512, 2048, 8192},
                xticklabels={32, 128, 512, 2k, 8k},
                legend style={at={(0.98,0.98)}, anchor=north east, font=\tiny\fontfamily{qag}\selectfont, draw=SoftGridColor, row sep=0.1pt},
                axis line style={AxisFrameColor}
            ]
                \addplot[only marks, mark=*, mark size=1.25pt, DataPointColor] 
                    table {plot_data/dim_only_hero/msmarco_dev_bert_l8.dat};
                \addplot[FitLineColor, very thick, dashed, domain=32:8192, samples=100] 
                    { (\EqA / x)^(\EqAlpha) + \EqE};
            \end{axis}
        \end{tikzpicture}
    \end{subfigure}%
    \begin{subfigure}[b]{0.245\textwidth}
        \centering
        \begin{tikzpicture}
            \newcommand{\EqA}{30.41925181238873}
            \newcommand{\EqAlpha}{1.4838412883235625}
            \newcommand{\EqE}{0.34120542124708536}
            
            \begin{axis}[
                title={\small BERT - TREC DL},
                xmode=log, log basis x={2},
                grid=major,
                scale only axis,
                grid style={solid, SoftGridColor},
                width=0.75\textwidth, %
                height=2.0cm,
                tick pos=left, 
                tick label style={font=\footnotesize},
                title style={font=\footnotesize\scshape, yshift=-0.75ex},
                y tick label style={/pgf/number format/fixed, font=\footnotesize, /pgf/number format/assume math mode=true},
                scaled y ticks=false,
                xtick={32, 128, 512, 2048, 8192},
                xticklabels={32, 128, 512, 2k, 8k},
                legend style={at={(0.98,0.98)}, anchor=north east, font=\tiny\sffamily, draw=SoftGridColor, row sep=0.1pt},
                axis line style={AxisFrameColor}
            ]
                \addplot[only marks, mark=*, mark size=1.25pt, DataPointColor] 
                    table {plot_data/dim_only_hero/trec_dl_bert_l8.dat};
                \addplot[FitLineColor, very thick, dashed, domain=32:8192, samples=100] 
                    { (\EqA / x)^(\EqAlpha) + \EqE};
            \end{axis}
        \end{tikzpicture}
    \end{subfigure}%
    \begin{subfigure}[b]{0.245\textwidth}
        \centering
        \begin{tikzpicture}
            \newcommand{\EqA}{1.8514831857373735}
            \newcommand{\EqAlpha}{0.9110743898928739}
            \newcommand{\EqE}{0.05807020163219602}
            
            \begin{axis}[
                title={\small Ettin - MSMARCO Dev},
                xmode=log, log basis x={2},
                scale only axis,
                grid=major,
                grid style={solid, SoftGridColor},
                width=0.75\textwidth,
                height=2.0cm,
                tick pos=left,
                tick label style={font=\footnotesize},
                title style={font=\footnotesize\scshape, yshift=-0.75ex},
                y tick label style={/pgf/number format/fixed, font=\footnotesize, /pgf/number format/assume math mode=true},
                scaled y ticks=false,
                y tick label style={/pgf/number format/fixed},
                xtick={128, 512, 2048, 8192},
                xticklabels={128, 512, 2k, 8k},
                legend style={at={(0.98,0.98)}, anchor=north east, font=\tiny\sffamily, draw=SoftGridColor},
                axis line style={AxisFrameColor}
            ]
                \addplot[only marks, mark=*, mark size=1.25pt, DataPointColor] 
                    table {plot_data/dim_only_hero/msmarco_dev_ettin_l19.dat};
                \addplot[FitLineColor, very thick, dashed, domain=128:8192, samples=100] 
                    { (\EqA / x)^(\EqAlpha) + \EqE};
            \end{axis}
        \end{tikzpicture}
    \end{subfigure}%
    \begin{subfigure}[b]{0.245\textwidth}
        \centering
        \begin{tikzpicture}
            \newcommand{\EqA}{10.911776456614735}
            \newcommand{\EqAlpha}{0.7494677583912737}
            \newcommand{\EqE}{0.6900086441812947}
            
            \begin{axis}[
                title={\small Ettin - TREC DL},
                xmode=log, log basis x={2},
                scale only axis,
                grid=major,
                grid style={solid, SoftGridColor},
                width=0.75\textwidth,
                height=2.0cm,
                tick pos=left,
                tick label style={font=\footnotesize},
                title style={font=\footnotesize\scshape, yshift=-0.75ex},
                y tick label style={/pgf/number format/fixed, font=\footnotesize, /pgf/number format/assume math mode=true},
                scaled y ticks=false,
                y tick label style={/pgf/number format/fixed},
                xtick={32, 128, 512, 2048, 8192},
                xticklabels={32, 128, 512, 2k, 8k},
                legend style={at={(0.98,0.98)}, anchor=north east, font=\tiny\sffamily, draw=SoftGridColor},
                axis line style={AxisFrameColor}
            ]
                \addplot[only marks, mark=*, mark size=1.25pt, DataPointColor] 
                    table {plot_data/dim_only_hero/trec_dl_ettin_l19.dat};
                \addplot[FitLineColor, very thick, dashed, domain=128:8192, samples=100] 
                    { (\EqA / x)^(\EqAlpha) + \EqE};
            \end{axis}
        \end{tikzpicture}
    \end{subfigure}

    \vspace{2pt}
    {\small\sffamily Embedding Dimension}

    \caption{Scaling behavior of contrastive entropy relative to embedding dimension on TREC DL Combined and MSMARCO Dev. The points represent the observed contrastive entropy while the line represents the fitted dimension-only scaling law. The BERT model shown in both plots is BERT-L8-H512-A8. The Ettin model shown in both plots is Ettin L19-H512-A8.}
    \label{fig:dim_only_hero}
\end{figure*}
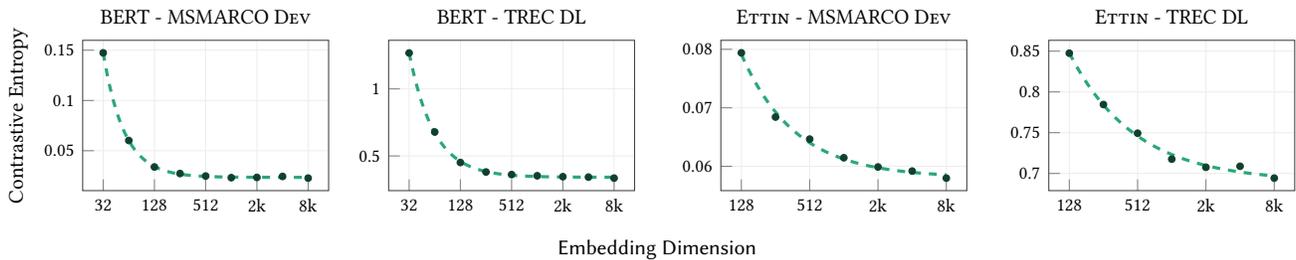

\section{Introduction}

Dense retrieval has emerged as a cornerstone of modern neural information retrieval (IR) due to its effectiveness and simplicity. By encoding queries and documents into a shared latent space, bi-encoder architectures \citep{dense_passage_retrieval, SNRM} enable efficient retrieval over massive corpora using maximum inner product search (MIPS). However, the scope of dense retrieval has recently expanded beyond simple topical matching to include more sophisticated tasks such as instruction following and multi-step reasoning. As these tasks grow in complexity, the inherent limitations of the underlying data structure, i.e., fixed-dimensional vectors, become of increasing importance.

Previous work has shown that the capacity of a dense retriever is fundamentally constrained by the geometry of its embedding space. Theoretical analyses have demonstrated that the ability of a model to perfectly separate relevant documents from a non-relevant set is strictly bounded by the embedding dimension $d$ \citep{hypencoder_hypernetworks_retrieval, theoretical_limitations_embedding_retrieval}. For a corpus of size $M > d + 1$, there exist relevance patterns that cannot be represented by any linear similarity function. Despite these theoretical insights, the empirical relationship between embedding dimensionality and retrieval effectiveness has not been systematically quantified in the same way that model parameters or training data have been characterized through scaling laws \citep{scaling_laws_neural_language_models, scaling_laws_dense_retrieval}.

Though exploring the scaling behavior of embedding dimensions provides useful insight to existing theories, understanding how retrieval performance scales with embedding dimension is not just a theoretical exercise; it is a practical tool for system design as demonstrated by the impact of prior scaling laws \cite{chinchilla_compute_optimal_llms, scaling_laws_neural_language_models}. As an example of the impact scaling laws can have, the Chinchilla scaling laws proposed by \citet{chinchilla_compute_optimal_llms} showed that for a limited compute budget an optimally trained model has both compute and data scaled equally. This finding allowed the Chinchilla model to outperform the existing Gopher model with ~20\% of the parameters and broadly empowered more efficient large language model training. Additionally, scaling laws enable experiments to be done at a small scale while being able to directly predict the outcome on a scaled up version.

As shown by the impact of prior scaling laws, being able to understand and predict how different parameters impact performance is an essential tool for building effective and efficient models. For dense retrieval efficiency, the choice of embedding dimension $d$ directly dictates the storage footprint of the document index and the computational cost of retrieval. These concerns are especially acute for on-device settings where disk space is often at a premium. Despite the impact of embedding dimension on efficiency, practitioners often rely on the ``native'' hidden size of a transformer backbone (e.g., 768 for BERT-base \cite{bert_bidirectional_transformers}), though it is unclear whether this choice is optimal or how performance is impacted as dimensions are expanded or compressed. Furthermore, the interplay between model size (parameters) and embedding size (representation capacity) remains poorly understood, leaving architects without a clear guide for balancing the two under fixed compute or storage budgets. 

In this work, we conduct a comprehensive empirical study of embedding dimension scaling across two distinct model families: the classic BERT \cite{bert_bidirectional_transformers} architecture and the more recent Ettin suite \cite{Ettin_suite}. By training models across a wide range of parameter counts and embedding dimensions, including dimensions that exceed the backbone's hidden size, we provide a detailed picture of retrieval scaling behavior. Our analysis reveals that retrieval performance, measured by contrastive entropy, follows a predictable power law relative to the embedding dimension. We further derive a joint scaling law that accounts for both model size and dimensionality, offering a unified framework for predicting performance.

Our findings show that for retrieval tasks aligned with the training objective, increasing the embedding dimension yields consistent, albeit diminishing, returns. However, we observe that this scaling behavior is less consistent for out-of-domain tasks which differ from the training distribution though generally increasing the embedding dimension still improves performance. Finally, we demonstrate the utility of our scaling laws by providing a cost-aware analysis, showing the optimal model and embedding size to maximize retrieval effectiveness under specific inference compute constraints. We hope our work serves as a practical guide for the development of next-generation retrieval systems that are both effective and efficient.

\vspace{-5pt}
\section{Related Work}

\input{plot_sources/bert_joint_scaling_law_figure}
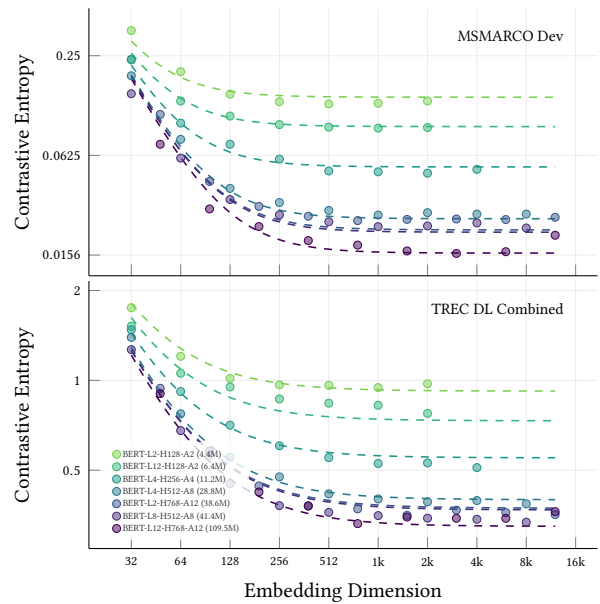
\begin{figure}[t]
\centering
\definecolor{viridisettinL7H256A4}{rgb}{0.478, 0.821, 0.318}
\definecolor{viridisettinL10H384A6}{rgb}{0.171, 0.694, 0.494}
\definecolor{viridisettinL19H512A8}{rgb}{0.135, 0.545, 0.554}
\definecolor{viridisettinL22H768A12}{rgb}{0.198, 0.392, 0.555}
\definecolor{viridisettinL28H1024A16}{rgb}{0.271, 0.214, 0.507}
\definecolor{viridisettinL28H1792A28}{rgb}{0.267, 0.005, 0.329}
\begin{tikzpicture}
\begin{groupplot}[
    group style={
        group size=1 by 2,
        xlabels at=edge bottom,
        ylabels at=edge left,
        x descriptions at=edge bottom,
        vertical sep=2pt,
    },
    width=0.38\textwidth, height=0.2\textwidth,
    xmode=log, ymode=log,
    ylabel absolute,
    scale only axis,
    log basis x={2}, log basis y={2},
    tick label style={font=\tiny},
    ylabel style={at={(0.09, 0.5)}, anchor=south},
    xlabel={\small Embedding Dimension}, ylabel={\small Contrastive Entropy},
    grid=both, grid style={line width=.1pt, draw=gray!15},
    axis lines*=left,
    xtick={64,128,256,512,1024,2048,4096,8192,16384,32768},
    xticklabels={64,128,256,512,1k,2k,4k,8k,16k,32k},
    log ticks with fixed point,
    yticklabel style={/pgf/number format/fixed, /pgf/number format/precision=1},
]
    \nextgroupplot[
]
        \node[anchor=north east, font=\scriptsize] at (rel axis cs:0.95,0.95) {MSMARCO Dev};
        \addplot[no marks, viridisettinL7H256A4, dashed, semithick, domain=64:28672, samples=40, forget plot] {0.017873426099070767 + 1.2662719509189801 * (16.79744^(-0.8053334325261825)) + 3.135009519918703 * (x^(-1.0151223099440774))};
        \addplot[only marks, mark=*, mark size=1.5pt, viridisettinL7H256A4, fill opacity=0.5] coordinates {(64,0.19444143662004676) (128,0.16845574407327057) (256,0.16435574268894898) (512,0.1522827686776395) (1024,0.15044735710871873) (2048,0.15212362441981891) (4096,0.14963506375512767)};
        \addplot[no marks, viridisettinL10H384A6, dashed, semithick, domain=64:28672, samples=40, forget plot] {0.017873426099070767 + 1.2662719509189801 * (31.883136^(-0.8053334325261825)) + 3.135009519918703 * (x^(-1.0151223099440774))};
        \addplot[only marks, mark=*, mark size=1.5pt, viridisettinL10H384A6, fill opacity=0.5] coordinates {(96,0.12993819182065716) (192,0.10570008916631585) (384,0.10383172473172106) (768,0.0976834382865721) (1536,0.09657377909900612) (3072,0.09779274985376085) (6144,0.09637116876078104)};
        \addplot[no marks, viridisettinL19H512A8, dashed, semithick, domain=64:28672, samples=40, forget plot] {0.017873426099070767 + 1.2662719509189801 * (68.14464^(-0.8053334325261825)) + 3.135009519918703 * (x^(-1.0151223099440774))};
        \addplot[only marks, mark=*, mark size=1.5pt, viridisettinL19H512A8, fill opacity=0.5] coordinates {(128,0.07939574900196374) (256,0.0684215041985908) (512,0.0646450996753383) (1024,0.06146095012728942) (2048,0.059885345486817655) (4096,0.05918511865899575) (8192,0.05798212717667438)};
        \addplot[no marks, viridisettinL22H768A12, dashed, semithick, domain=64:28672, samples=40, forget plot] {0.017873426099070767 + 1.2662719509189801 * (149.014272^(-0.8053334325261825)) + 3.135009519918703 * (x^(-1.0151223099440774))};
        \addplot[only marks, mark=*, mark size=1.5pt, viridisettinL22H768A12, fill opacity=0.5] coordinates {(192,0.06045111564147056) (384,0.052766501151237044) (768,0.04965967297530983) (1536,0.04603157621410081) (3072,0.04477429470906437) (6144,0.04244246658085818) (12288,0.044093281237422385)};
        \addplot[no marks, viridisettinL28H1024A16, dashed, semithick, domain=64:28672, samples=40, forget plot] {0.017873426099070767 + 1.2662719509189801 * (394.781696^(-0.8053334325261825)) + 3.135009519918703 * (x^(-1.0151223099440774))};
        \addplot[only marks, mark=*, mark size=1.5pt, viridisettinL28H1024A16, fill opacity=0.5] coordinates {(256,0.03842915044608216) (512,0.03577413609432446) (1024,0.0307974109662835) (2048,0.027795699233451727) (4096,0.025093739337760726) (8192,0.025569632938856782) (16384,0.026518885155207082)};
        \addplot[no marks, viridisettinL28H1792A28, dashed, semithick, domain=64:28672, samples=40, forget plot] {0.017873426099070767 + 1.2662719509189801 * (1028.050688^(-0.8053334325261825)) + 3.135009519918703 * (x^(-1.0151223099440774))};
        \addplot[only marks, mark=*, mark size=1.5pt, viridisettinL28H1792A28, fill opacity=0.5] coordinates {(448,0.02529397555211321) (896,0.026192131249347557) (1792,0.023124561316011004) (3584,0.02248143674110241) (7168,0.024476695704646476) (14336,0.023093255144697004) (28672,0.022447750703329536)};
    \nextgroupplot[
        legend style={at={(0.03,0.03)}, anchor=south west, nodes={scale=0.4, transform shape}, fill opacity=0.8, draw=none},
        legend cell align={left}]
        \node[anchor=north east, font=\scriptsize] at (rel axis cs:0.95,0.95) {TREC DL Combined};
        \addplot[no marks, viridisettinL7H256A4, dashed, semithick, domain=64:28672, samples=40, forget plot] {0.45048937139134976 + 2.4444943692197105 * (16.79744^(-0.5888975540185005)) + 2.010894335730228 * (x^(-0.5051020708687886))};
        \addplot[only marks, mark=*, mark size=1.5pt, viridisettinL7H256A4, fill opacity=0.5] coordinates {(64,1.1802773484041618) (128,1.100113618477032) (256,1.0159400738254813) (512,1.0154533070266778) (1024,0.998136638373752) (2048,0.9780078371789548) (4096,0.9858145479255062)};
        \addlegendentry{Ettin-L7-H256-A4 (16.8M)}
        \addplot[no marks, viridisettinL10H384A6, dashed, semithick, domain=64:28672, samples=40, forget plot] {0.45048937139134976 + 2.4444943692197105 * (31.883136^(-0.5888975540185005)) + 2.010894335730228 * (x^(-0.5051020708687886))};
        \addplot[only marks, mark=*, mark size=1.5pt, viridisettinL10H384A6, fill opacity=0.5] coordinates {(96,0.9113136952213252) (192,0.8690191099176917) (384,0.8152838135488648) (768,0.7947872301751747) (1536,0.7818858778124732) (3072,0.7852851473057352) (6144,0.7790859466674769)};
        \addlegendentry{Ettin-L10-H384-A6 (31.9M)}
        \addplot[no marks, viridisettinL19H512A8, dashed, semithick, domain=64:28672, samples=40, forget plot] {0.45048937139134976 + 2.4444943692197105 * (68.14464^(-0.5888975540185005)) + 2.010894335730228 * (x^(-0.5051020708687886))};
        \addplot[only marks, mark=*, mark size=1.5pt, viridisettinL19H512A8, fill opacity=0.5] coordinates {(128,0.8474923156937858) (256,0.7845347757787893) (512,0.7493467474534402) (1024,0.7175804755491234) (2048,0.707602740436951) (4096,0.7088177816273029) (8192,0.6943004360586047)};
        \addlegendentry{Ettin-L19-H512-A8 (68.1M)}
        \addplot[no marks, viridisettinL22H768A12, dashed, semithick, domain=64:28672, samples=40, forget plot] {0.45048937139134976 + 2.4444943692197105 * (149.014272^(-0.5888975540185005)) + 2.010894335730228 * (x^(-0.5051020708687886))};
        \addplot[only marks, mark=*, mark size=1.5pt, viridisettinL22H768A12, fill opacity=0.5] coordinates {(192,0.7631846205074662) (384,0.7038415061315808) (768,0.6486690161664599) (1536,0.6238172114077761) (3072,0.6163331768136974) (6144,0.6107320932099604) (12288,0.6058306483282715)};
        \addlegendentry{Ettin-L22-H768-A12 (149.0M)}
        \addplot[no marks, viridisettinL28H1024A16, dashed, semithick, domain=64:28672, samples=40, forget plot] {0.45048937139134976 + 2.4444943692197105 * (394.781696^(-0.5888975540185005)) + 2.010894335730228 * (x^(-0.5051020708687886))};
        \addplot[only marks, mark=*, mark size=1.5pt, viridisettinL28H1024A16, fill opacity=0.5] coordinates {(256,0.6804524230802926) (512,0.6140188367190038) (1024,0.5979844774041548) (2048,0.5772210562240848) (4096,0.5788906119046824) (8192,0.5650053634724014) (16384,0.5510452281127601)};
        \addlegendentry{Ettin-L28-H1024-A16 (394.8M)}
        \addplot[no marks, viridisettinL28H1792A28, dashed, semithick, domain=64:28672, samples=40, forget plot] {0.45048937139134976 + 2.4444943692197105 * (1028.050688^(-0.5888975540185005)) + 2.010894335730228 * (x^(-0.5051020708687886))};
        \addplot[only marks, mark=*, mark size=1.5pt, viridisettinL28H1792A28, fill opacity=0.5] coordinates {(448,0.5935451752911703) (896,0.5474052538482646) (1792,0.4843536354681664) (3584,0.5039361652811687) (7168,0.48027273888411304) (14336,0.4921053168396172) (28672,0.4731240147424193)};
        \addlegendentry{Ettin-L28-H1792-A28 (1028.1M)}
\end{groupplot}
\end{tikzpicture}
\caption{Empirical results and joint scaling laws for the Ettin model family on MSMARCO Dev and TREC DL Combined. The points represent empirical results at various embedding dimensions and model sizes which are represented by the point colors. The dashed lines represent the joint scaling laws fit on the observed data.}
\label{fig:ettin_joint_scaling_plot}
\end{figure}

\paragraph{\textbf{Theoretical Impact of Embedding Dimension}}
Theoretical analysis highlights the geometric bottlenecks of single-vector dense retrieval, where the capacity to model relevance is fundamentally tied to dimensionality \citep{sparse_dense_attentional_representations, defense_dual_encoders_ranking}. \citet{hypencoder_hypernetworks_retrieval} apply Radon's theorem to demonstrate that fixed-dimensional spaces cannot linearly separate arbitrary relevant sets once the corpus size exceeds the dimension plus one. Similarly, \citet{theoretical_limitations_embedding_retrieval} prove that the number of realizable top-$k$ result sets is strictly bounded by dimensionality, a limitation which they show can happen even with simple natural language queries. These findings suggest an inherent expressivity ceiling for inner-product dense retrieval that is tied to the embedding dimension, which motivates our systematic evaluation of how embedding dimension dictates performance.

\paragraph{\textbf{Dense Retrieval and Efficient Representations}}
Dense retrieval models, such as DPR \citep{dense_passage_retrieval}, encode text into fixed vectors, with effectiveness significantly improved via hard negative selection \cite{ance_negative_contrastive_learning, rance, rocketqav2_joint_training}, distillation \cite{cross_architecture_knowledge_distillation}, and self-supervised pretraining \cite{condenser_pretraining_dense_retrieval, contriever_unsupervised_contrastive_learning}. However, the embedding dimension---a primary driver of storage and latency for both exhaustive and approximate nearest neighbor (ANN) systems like Faiss \citep{faiss_library}---is typically treated as a static parameter derived from the encoder's hidden size. 
There has been some work on minimizing the size of the embedding dimension via compression or dimensionality reduction \citep{redundancy_elimination_dense_vectors, sparse_dense_attentional_representations}, but these approaches generally operate within or below the bounds of the backbone's hidden dimension. Matryoshka Representations \citep{matryoshka_representations_adaptive} are similar, but allow for embedding truncation after encoding. Other approaches like product quantization \citep{product_quantization_nearest_neighbor} can help reduce embedding space as well. Our work is complementary to many of the explicit compression techniques and efficient search techniques which reduce the compute cost to retrieve as we consider possibly reducing the dimension of the embeddings for added efficiency. In contrast to existing approaches to reduce the embedding dimension, we treat the target embedding dimension as a controlled variable and systematically study both compression and expansion relative to the encoder hidden size, including regimes larger than the backbone. Our comprehensive analysis allows us to derive scaling laws that accurately predict the retrieval performance given different embedding dimensions which has not been possible before this point.

\paragraph{\textbf{Scaling Laws for Retrieval}}
Scaling laws have successfully characterized performance relative to parameters, data, and compute in language modeling \citep{scaling_laws_neural_language_models, chinchilla_compute_optimal_llms} and dense retrieval \citep{scaling_laws_dense_retrieval, retrieval_capabilities_scaling_flops}.  \citeauthor{scaling_sparse_dense_decoder_llms} expanded dense only scaling behavior to compared scaling trends across sparse versus dense paradigms. Though scaling laws exist for dense retrieval, all prior work treats the embedding dimension as a fixed hyperparameter. In this work we specifically isolate the embedding dimension to understand how it contributes to retrieval performance as well as deriving a joint scaling law for embedding dimension and model size.

\section{Methodology}
In this section, we describe the experimental setup we use to understand the scaling behavior of embedding dimensions. Our setup is largely similar to prior work in terms of model architecture and training approach to make our findings as realistic and directly applicable as possible. The few times we diverge from common training and architectural approaches are in service of our core research question, which is to understand how embedding scaling impacts retrieval performance.

\subsubsection{Base Encoders}
\begin{table*}[htbp]
\centering
\caption{Information about the models used in our experiments and the parameters of fitting our dimension only scaling law on MSMARCO Dev and TREC DL Combined.}
\label{tab:model_details_and_dim_scaling_laws}
\scalebox{0.9}{
\begin{tabular}{l !{\color{lightgray}\vrule} c c !{\color{lightgray}\vrule} cccc !{\color{lightgray}\vrule} cccc}
\toprule
\multicolumn{1}{l}{} & \multicolumn{1}{c}{} & \multicolumn{1}{c}{} & \multicolumn{4}{c}{\textbf{MSMARCO Dev}} & \multicolumn{4}{c}{\textbf{TREC DL}} \\
\multicolumn{1}{l}{\textbf{Model Name}} & \multicolumn{1}{c}{\textbf{Size}} & \multicolumn{1}{c}{\textbf{Embed Dims}} & \multicolumn{1}{c}{$A$} & \multicolumn{1}{c}{$\alpha$} & \multicolumn{1}{c}{$\delta_D$} & \multicolumn{1}{c}{$R^2$} & \multicolumn{1}{c}{$A$} & \multicolumn{1}{c}{$\alpha$} & \multicolumn{1}{c}{$\delta_D$} & \multicolumn{1}{c}{$R^2$} \\
\midrule
BERT-L2-H128-A2 & 4.39 M & 32-2048 & 84.73 & 1.71 & 0.128 & 0.9988 & 319.19 & 1.73 & 0.953 & 0.9984 \\
BERT-L12-H128-A2 & 6.37 M & 32-2048 & 61.16 & 1.75 & 0.092 & 0.9991 & 72.49 & 1.34 & 0.810 & 0.9905 \\
BERT-L4-H256-A4 & 11.17 M & 32-4096 & 117.48 & 1.86 & 0.052 & 0.9962 & 63.14 & 1.21 & 0.517 & 0.9993 \\
BERT-L4-H512-A8 & 28.76 M & 32-8192 & 62.52 & 1.72 & 0.027 & 0.9995 & 105.71 & 1.35 & 0.394 & 0.9992 \\
BERT-L2-H768-A12 & 38.60 M & 48-12288 & 472.56 & 2.23 & 0.026 & 0.9996 & 179.78 & 1.48 & 0.359 & 0.9989 \\
BERT-L8-H512-A8 & 41.37 M & 32-8192 & 55.21 & 1.76 & 0.024 & 0.9998 & 158.77 & 1.48 & 0.341 & 0.9997 \\
BERT-L12-H768-A12 & 109.48 M & 48-12288 & 170.38 & 2.08 & 0.018 & 0.9917 & 96.42 & 1.33 & 0.342 & 0.9953 \\
\midrule
Ettin-L7-H256-A4 & 16.80 M & 64-4096 & 3.46 & 1.05 & 0.149 & 0.9771 & 9.04 & 0.91 & 0.977 & 0.9782 \\
Ettin-L10-H384-A6 & 31.88 M & 96-6144 & 58.68 & 1.64 & 0.097 & 0.9806 & 4.47 & 0.75 & 0.769 & 0.9802 \\
Ettin-L19-H512-A8 & 68.14 M & 128-8192 & 1.75 & 0.91 & 0.058 & 0.9953 & 6.00 & 0.75 & 0.690 & 0.9941 \\
Ettin-L22-H768-A12 & 149.01 M & 192-12288 & 0.83 & 0.73 & 0.042 & 0.9857 & 11.27 & 0.80 & 0.598 & 0.9923 \\
Ettin-L28-H1024-A16 & 394.78 M & 256-16384 & 0.38 & 0.58 & 0.024 & 0.9467 & 14.79 & 0.87 & 0.558 & 0.9709 \\
Ettin-L28-H1792-A28 & 1028.05 M & 448-28672 & 0.06 & 0.45 & 0.022 & 0.5325 & 72.94 & 1.05 & 0.479 & 0.9187 \\
\bottomrule
\end{tabular}}
\end{table*}

To get a complete picture of the scaling behavior of embedding dimensions, we use two encoder model families, BERT \cite{bert_bidirectional_transformers} and Ettin \cite{Ettin_suite}. BERT represents one of the most popular encoder models while Ettin is a more recent suite and thus adopts many advancements in model architecture and training recipe. Both model families include a range of models trained on the same data making comparisons between models of different sizes possible. As model size plays a crucial role in both computational cost and retrieval performance, we select several model sizes across both families to understand the interplay between scaling model size and scaling embedding size. We select model sizes to get a diverse range of sizes and configurations. The specific models used along with their parameter counts are presented in Table \ref{tab:model_details_and_dim_scaling_laws}. 

We use two distinct architectures to ensure that our results generalize beyond a specific model configuration. To further bolster our results, each architecture uses a different training recipe based on popular dense retrieval approaches. We believe this is an important choice to understand the generalization of our findings and enables a more complete analysis of how different training choices impact scaling behavior.

\subsection{Model Architecture}
Following existing dense retrieval approaches, we use a two-tower approach which uses a shared transformer encoder $E$ to encode both the queries and documents \cite{attention_is_all_you_need}. Formally, given text $T$ with tokens $t_1, t_2, \dots, t_n$, we use $E$ to encode the text into hidden states $h_1, h_2, \dots, h_n$ where $h_i \in \mathbb{R}^{d}$ and $d$ is the hidden size of $E$. We then scale these values using a linear layer:
$$
h'_i = Wh_i + b
$$
where $W \in \mathbb{R}^{m \times d}$ represents the weight matrix and $b \in \mathbb{R}^{m}$ represents the bias vector where $m$ represents the final embedding dimension. To get the final embedding $e$, we take the element-wise mean over all the hidden dimensions in the sequence.
$$
e = \frac{1}{n} \sum_{i=1}^n h'_i
$$
We selected mean pooling over other pooling methods as it has a smaller bottleneck than selection based pooling methods such as selecting the start token or end token. For the Ettin models, we normalize the final embeddings and apply a fixed temperature of $\tau=0.02$ as larger hidden embedding dimensions used by Ettin can cause unnormalized dot-product similarities to grow with dimension, leading to inflated similarity values and numerical instability.

\subsection{Training Data}
To fully utilize the underlying architecture of the model families and generalize our results beyond specific training data, we use two different training datasets for the two different encoder families. For the BERT-based models, we use the MSMARCO Passage dataset \citep{ms_marco_dataset} for training. Specifically, the version preprocessed with teacher scores from \citet{hypencoder_hypernetworks_retrieval}. The dataset contains approximately 500k training queries. Each training instance consists of: (1) a query (2) one or more known relevant passage (from MSMARCO annotations) (3) Around 200 additional candidate passages. All query-passage pairs include a score from the cross-encoder model \path{cross-encoder/ms-marco-MiniLM-L-12-v2}  to provide teacher scores for training. 

For the Ettin models, we use MSMARCO Instruct \cite{ms_marco_dataset}. This dataset augments the standard MSMARCO Passage dataset with LLM generated instructions to the original queries. Additional hard negatives are also generated to encourage instruction following. As the instruction-augmented queries are in addition to the existing MSMARCO queries this dataset has approximately 1M training queries. We selected this dataset as prior work suggested it produced robust performance on complex retrieval tasks \cite{CRUMB} and we wanted to understand if longer, more complex queries would have an impact on the scaling behavior.

During training, we sample one positive and seven negative passages per query. For the BERT-based model we selected the most relevant passage as judged by the teacher as the positive. For the MSMARCO Instruct data we prioritize sampling the up to three negatives which were generated to ensure the models are able to understand instructions, and the rest are sampled from query negatives. It is worth noting that for the BERT data, although we call the passages negative following prior work, the passages might be relevant as they are sampled from retrieved documents for the query.

\subsection{Training Objective}
Our primary training recipe for the BERT models combines two loss functions, a knowledge distillation and contrastive learning loss. We made this choice to reflect current popular training practices which often combine the two for improved performance \cite{hypencoder_hypernetworks_retrieval, scaling_sparse_dense_decoder_llms, ColBERTv2, splade_v3}. For MSMARCO Instruct, we only use contrastive learning loss as the dataset does not have labels and standard cross-encoders may not be well suited for the instruction following task. Additionally, this is also a common training setup for dense retrieval, so it is worth representing in our experiments.

\subsubsection{Margin Mean Squared Error}
The Margin MSE loss \cite{Margin_MSE_loss} encourages the encoder to reproduce the differences in score margin  from a teacher cross-encoder. For a query $q$, positive passage embedding $p^+$, and negative passage embedding $p^-$, the loss is:
\begin{equation}
    \mathcal{L}_{\text{margin-mse}} = (\Delta_{\text{student}} - \Delta_{\text{teacher}})^2
\end{equation}
where $\Delta_{\text{student}} = (e_q \cdot e_{p^+}) - (e_q \cdot e_{p^-})$ and $\Delta_{\text{teacher}} = s_{\text{teacher}}(q, p^+) - s_{\text{teacher}}(q, p^-)$. The values $e_q$, $e_{p^+}$, and $e_{p^-}$ represent the encoded query, positive passage, and negative passage, respectively, and $s_{\text{teacher}}$ denotes the teacher cross-encoder scores. We apply MarginMSE to each query and hard negative pair in the batch and take the mean to get the final loss.

\subsubsection{Cross-Entropy Contrastive Loss}
The contrastive loss treats retrieval as a classification problem over passages in each batch. For a query $q$ with positive passage $p^+$ and negative passages $\mathcal{N}$ that vary for the BERT training and Ettin training the contrastive loss is defined as:
\begin{equation}
    \mathcal{L}_{\text{contrastive}} = -\log\frac{\exp(e_q \cdot e_{p^+})}{\exp(e_q \cdot e_{p^+}) + \sum_{p^- \in \mathcal{N}}\exp(e_q \cdot e_{p^-})}
\end{equation}
For BERT, we do not include the seven hard negatives in $\mathcal{N}$ as the sampling strategy results in relevant passages in the hard negatives, instead we use only positive passages for the other in-batch queries. For Ettin, we use all in-batch passages except $p^+$. Additionally, for the Ettin models we use a fixed temperature of $\tau=0.02$ to account for the limited range due to the embedding normalization.

\subsubsection{Combined Objective}
The final training loss for BERT is:
\begin{equation}
    \mathcal{L}_{\text{total}} = \mathcal{L}_{\text{margin-mse}} + \mathcal{L}_{\text{contrastive}}
\end{equation}
while for Ettin, it is:
\begin{equation}
    \mathcal{L}_{\text{total}} = \mathcal{L}_{\text{contrastive}}
\end{equation}
The choice of two losses allows us to better understand the impact of training on down-stream retrieval performance.

\subsection{Training Configuration}
As the focus of our experiments are on understanding the behavior of embedding and model scaling we use a fixed number of training epochs for all models in the same model family. For the BERT models, we use 50 epochs, while for the Ettin models we use 15 as they have more training data per epoch. We use PyTorch \cite{PyTorch} and Huggingface \cite{HuggingFace} for our implementation. We use AdamW as our optimizer with a learning rate of 2e-5, and a linear learning rate scheduler with a warm up ratio of 0.05. We use an effective batch size of 128 for BERT training and an effective batch size of 64 for Ettin training. For all other hyperparameters, we use the default parameters from the Huggingface trainer.

A crucial choice for our analysis is which set of final embedding dimensions $m$ to select. We decided to use multiples of each encoder's native hidden size; for all models we used the multiples $\Phi: \{\frac{1}{4}, \frac{1}{2}, 1, 2, 4, 8, 16 \}$ so that for a model with hidden size $d$ the set of embedding dimensions is $\{\phi\cdot d: \phi \in \Phi\}$. For some of the larger BERT models, we included some additional lower multiples $\{\frac{1}{16}, \frac{1}{8}\}$ to understand how they behaved with additional compression.

\subsection{Evaluation Data}
Following prior work on scaling laws, we use evaluation data that is closely aligned with the training data \cite{scaling_laws_dense_retrieval, scaling_laws_neural_language_models, chinchilla_compute_optimal_llms}. Specifically, we use a combined version of TREC Deep Learning (DL) 2019, 2020, and 2021 \cite{trec_2020_deep_learning, trec_dl_21, trec_dl_2020} and MSMARCO Dev \cite{ms_marco_dataset} for our main analysis. Both of these datasets include similar data to the training data with both featuring an ad-hoc retrieval task with natural language queries. MSMARCO Dev contains approximately 7k queries with incomplete labels that include one to two relevant passage per query. Our combined TREC DL contains 150 queries with far more complete labels and graded relevance labels, constructed through pooling.

We use these two datasets for deriving scaling laws and our core analysis as using less aligned data adds additional confounding factors. With that said, understanding how performance on out-of-domain (i.e. unaligned) tasks is impacted by scaling is an important consideration. Thus, we include two additional evaluation tasks from the CRUMB benchmark \cite{CRUMB, doris_mae, legal_retrieval_benchmark} which have markedly different characteristics from the training data. The first task is Legal QA which has reasoning-intense legal questions as queries and legal statutes as documents. The second is Paper Retrieval where the queries are paragraphs describing features of a desired paper and the documents are paper titles and abstracts. We selected these tasks because the underlying tasks is still similar to the training task (i.e. retrieval), but the domain and query format are both different which should capture the models ability to generalize.

For TREC DL, MSMARCO Dev, and Legal QA we consider all passages in the QRELs with relevance above zero to be positive. For Paper Retrieval, we use the relevant passages provided by the binary QRELs.

\subsection{Evaluation Metrics}
\subsubsection{Metrics}
Following prior work investigating dense retrieval scaling laws \cite{scaling_laws_dense_retrieval}, our primary evaluation metric is contrastive entropy. Contrastive entropy is a smooth measure of a retrieval models ability to separate a relevant document from irrelevant documents. The implementation is nearly identical to the contrastive learning loss we use:
$$
L = -\log\frac{\exp(e_q \cdot e_{p^+})}{\exp(e_q \cdot e_{p^+}) + \sum_{p^- \in \mathcal{N}}\exp(e_q \cdot e_{p^-})}
$$
Where $e_q$, $e_{p^+}$, and $e_{p^-}$ represent the encoded query, positive passage, and negative passage, respectively. For the negatives, $\mathcal{N}$, following \citet{scaling_laws_dense_retrieval} we use 256 passages sampled uniformly at random from the entire corpus with the exception of known positive passages. For datasets that have multiple positives for a query, we calculate contrastive entropy for each query-positive pair. The final contrastive entropy was found by taking the mean for all positives for the same query and then the mean over these query values. Like the contrastive loss we use during training, for Ettin models we apply a temperature of $\tau=0.02$.

To select the best checkpoint, we follow the approach used by \citet{scaling_laws_dense_retrieval} and run contrastive entropy, using our evaluation datasets, on each model checkpoint and select the best value. As we are interested in the scaling behavior this is a valid approach to remove potential noise.

To verify that the observed scaling behavior meaningfully reflects retrieval performance under standard ranking metrics, we study the empirical scaling results of a subset of our evaluation configuration based on reciprocal rank (RR@10) and recall (R@1000) in section \ref{sec:ir_metric_scaling}.

\subsubsection{Why Contrastive Entropy?}
As we aim to study embedding dimension scaling and derive scaling laws based on empirical results, our evaluation metric needs to be continuous and able to measure minor retrieval performance improvements. Traditional ranking metrics are less well-suited for this purpose as they are inherently discrete and insensitive to improvements beyond their cutoff parameter, while sensitive to small changes in ranking order. As contrastive entropy provides a continuous measure of a model's ability to separate relevant from non-relevant documents and has been shown in prior work \cite{scaling_laws_dense_retrieval} to exhibit strong correlation with conventional retrieval metrics, we select it as our primary evaluation measure.

\begin{table}[htbp]
\centering
\caption{Joint scaling law parameters and goodness-of-fit ($R^2$) for BERT and Ettin model families on TREC DL combined and MSMARCO Dev.}
\label{tab:joint_scale_parameters}
\scalebox{0.9}{
\begin{tabular}{l !{\color{lightgray}\vrule} l !{\color{lightgray}\vrule} cccccc}
\toprule
\textbf{Model} & \textbf{Set} & $A$ & $B$ & $\alpha$ & $\beta$ & $\delta$ & $R^2$ \\
\midrule
\multirow{2}{*}{BERT}  & MS   & 114.887 & 0.800 & 1.887 & 1.247 & 0.013 & 0.975 \\
                       & TREC & 85.543 & 2.588 & 1.316 & 0.961 & 0.295 & 0.986 \\
\midrule
\multirow{2}{*}{Ettin} & MS   & 3.135 & 1.266 & 1.015 & 0.805 & 0.017 & 0.997 \\
                       & TREC & 2.010 & 2.444 & 0.505 & 0.588 & 0.450 & 0.978 \\
\bottomrule
\end{tabular}}
\end{table}

\section{Results and Analysis}
In this section, we present the results of our experiments and derive scaling laws based on our empirical results.

\subsection{Embedding Dimension Scaling Behavior}
Given a model, we want to understand how scaling the embedding dimension impacts contrastive entropy, and, by extension, retrieval metrics. We do this by scaling the embedding dimension over a fixed range. 

The results of this process are shown in Figure \ref{fig:dim_only_hero} for a selected BERT and Ettin model on our core evaluation tasks. The points, which represent the empirical results, follow a clear pattern; as embedding dimension increases, there is a decrease in contrastive entropy. Note that lower contrastive entropy is better and correlates to higher standard retrieval metrics. Although there is a clear decrease, as the embedding dimension continues to scale, the decrease begins to plateau. These results are interesting from a theoretical point of view as many theoretical limitations of embedding capacity are tied to the embedding dimension \cite{hypencoder_hypernetworks_retrieval,theoretical_limitations_embedding_retrieval} and one might expect that the gains in performance would continue with less saturation. From another theoretical perspective, the number of near orthogonal vectors increases exponentially with the number of dimensions \cite{nearly_orthogonal}, as near orthogonal vectors allow for more information with minimal interference, they are likely a good proxy for capacity. At some point, the amount of information necessary to do a retrieval task is achieved and any further gains are diminishing. Additionally, as text and likely the concepts and features within text, follow a Zipfian distribution \cite{zipfian_distribution} at a certain point the vast majority of likely features are covered and adding dimensions only helps for the infrequent long-tail features that have minimal impact on average evaluation performance.

The empirical results on their own are useful to better understand the impact of embedding dimension on retrieval performance, but in order to have a general sense of how scaling embeddings impacts performance it is useful to have an equation which can predict contrastive entropy given a model's embedding dimension. Inspired by classical risk decomposition, we find that the contrastive entropy parameterized by the embedding dimension is well approximated by:
\begin{equation} \label{eq:dim_only_scaling_law}
    L(D) =\frac{A}{D^{\alpha}} + \delta_D
\end{equation}
Where $L(D)$ is the predicted contrastive entropy and $D$ is the embedding dimension. The parameters $\delta_D$, $\alpha$, and $A$ are learned based on the empirical data. From a risk decomposition perspective, $\delta_D$ represents the irreducible error of the task which is inherent due to ambiguity in queries and potential missing or incorrect labels. Additionally, as we only include a term to represent the impact of the embedding dimension the irreducible error will also account for the model's limited ability to perfectly represent the task. We fit the empirical results to this equation using Scipy's $\texttt{curve\_fit}$ which does a non-linear least squares to fit the function.

The result of fitting Equation \eqref{eq:dim_only_scaling_law} to our empirical results is shown in Figure \ref{fig:dim_only_hero} with the dashed lines. The fitted parameters and $R^2$ value, which is a statistical measure of the proportion of the variance in a dependent variable that is explained by the independent variable, for all models on MSMARCO Dev and TREC DL Combined can be observed in Table \ref{tab:model_details_and_dim_scaling_laws}. The high values of $R^2$ mean that the variance in contrastive entropy is well explained by the embedding dimension; in other words, the predicted lines fit well. Nearly all models have $R^2 > 0.95$, but the largest Ettin on MSMARCO is quite a bit lower. This is likely because MSMARCO has lower contrastive entropy and so does the largest Ettin due to its size, this means noise is more likely to impact it and thus increase $R^2$. A similar pattern, though less pronounced, is also true for the other large Ettin models on both MSMARCO and TREC DL. Overall, the fit is remarkably good across model families, model sizes, and evaluation data. The empirical adherence to the derived scaling law suggests that the embedding dimension scaling behavior follows a power law. This finding is inline with prior work on scaling behavior of neural networks \cite{chinchilla_compute_optimal_llms, scaling_laws_dense_retrieval, scaling_laws_neural_language_models}, which also follow power laws. \textbf{Knowing that performance gains from embedding dimension scaling follows a precise power law provides a powerful tool for architects of future dense retrieval models which enables them to predict downstream performance given various choices for embedding dimension.} Additionally, there are applications to prunable embeddings such as Matryoshka representations \cite{matryoshka_representations_adaptive}, which allow embeddings to be pruned to any size after encoding. With our scaling laws, the embedding size to use at retrieval time could be dynamically adjusted with a clear understanding of how the choice will impact retrieval performance.

\begin{figure*}[t]
\centering
\definecolor{Set2bertL2H128A2}{rgb}{0.400, 0.761, 0.647}
\definecolor{Set2bertL12H128A2}{rgb}{0.988, 0.553, 0.384}
\definecolor{Set2bertL4H256A4}{rgb}{0.553, 0.627, 0.796}
\definecolor{Set2bertL4H512A8}{rgb}{0.906, 0.541, 0.765}
\definecolor{Set2bertL2H768A12}{rgb}{0.651, 0.847, 0.329}
\definecolor{Set2bertL8H512A8}{rgb}{1.000, 0.851, 0.184}
\definecolor{Set2bertL12H768A12}{rgb}{0.898, 0.769, 0.580}
\definecolor{Set2ettinL7H256A4}{rgb}{0.400, 0.761, 0.647}
\definecolor{Set2ettinL10H384A6}{rgb}{0.988, 0.553, 0.384}
\definecolor{Set2ettinL19H512A8}{rgb}{0.553, 0.627, 0.796}
\definecolor{Set2ettinL22H768A12}{rgb}{0.906, 0.541, 0.765}
\definecolor{Set2ettinL28H1024A16}{rgb}{0.651, 0.847, 0.329}
\definecolor{Set2ettinL28H1792A28}{rgb}{1.000, 0.851, 0.184}

\begin{tikzpicture}
\begin{groupplot}[
    group style={
        group size=2 by 2,
        horizontal sep=10pt, 
        vertical sep=30pt,
        xlabels at=edge bottom,
        ylabels at=edge left,
    },
    width=0.42\textwidth, height=0.2\textwidth, %
    xmode=log, ymode=log,
    ylabel absolute,
    ylabel style={at={(0.12, 0.5)}, anchor=south},
    log basis x={2}, log basis y={2},
    tick label style={font=\tiny},
    xlabel={\small Embedding Dimension}, ylabel={\small Contrastive Entropy},
    grid=both, grid style={line width=.1pt, draw=gray!15},
    axis lines*=left,
    xtick={32,64,128,256,512,1024,2048,4096,8192,16384,32768},
    xticklabels={32,64,128,256,512,1k,2k,4k,8k,16k,32k},
    log ticks with fixed point,
    yticklabel style={/pgf/number format/fixed, /pgf/number format/precision=1},
    legend style={
        nodes={scale=0.6, transform shape},
        draw=none, fill=none, legend columns=1,
        legend cell align={left},
        legend image code/.code={\draw[#1] plot coordinates {(0cm,0cm)};}
    }
]
    \nextgroupplot[
        title={\footnotesize BERT Family (Scientific Papers)},
        legend to name=legendBERT,
    ]
        \addplot[semithick, Set2bertL2H128A2, mark=*, mark size=1pt] coordinates {(32,4.4415) (64,3.5357) (128,3.5940) (256,3.2220) (512,3.2840) (1024,3.3623) (2048,3.4458)};
        \addlegendentry{BERT-L2-H128-A2 (4.4M)}
        \addplot[semithick, Set2bertL12H128A2, mark=*, mark size=1pt] coordinates {(32,4.1847) (64,3.4788) (128,3.3056) (256,3.4266) (512,3.4240) (1024,3.7735) (2048,3.8219)};
        \addlegendentry{BERT-L12-H128-A2 (6.4M)}
        \addplot[semithick, Set2bertL4H256A4, mark=*, mark size=1pt] coordinates {(32,4.0279) (64,2.9101) (128,2.6279) (256,2.4397) (512,2.3014) (1024,2.3638) (2048,2.4135) (4096,2.5300)};
        \addlegendentry{BERT-L4-H256-A4 (11.2M)}
        \addplot[semithick, Set2bertL4H512A8, mark=*, mark size=1pt] coordinates {(32,4.1535) (64,2.9950) (128,2.4080) (256,2.0038) (512,1.9534) (1024,1.9006) (2048,1.8748) (4096,1.8442) (8192,1.9490)};
        \addlegendentry{BERT-L4-H512-A8 (28.8M)}
        \addplot[semithick, Set2bertL2H768A12, mark=*, mark size=1pt] coordinates {(48,3.4082) (96,2.5469) (192,2.0870) (384,1.7755) (768,1.6437) (1536,1.6583) (3072,1.6868) (6144,1.6974) (12288,1.6127)};
        \addlegendentry{BERT-L2-H768-A12 (38.6M)}
        \addplot[semithick, Set2bertL8H512A8, mark=*, mark size=1pt] coordinates {(32,4.0453) (64,2.9589) (128,2.0679) (256,1.7609) (512,1.6904) (1024,1.6684) (2048,1.6697) (4096,1.7492) (8192,1.7688)};
        \addlegendentry{BERT-L8-H512-A8 (41.4M)}
        \addplot[semithick, Set2bertL12H768A12, mark=*, mark size=1pt] coordinates {(48,3.2604) (96,2.1560) (192,1.7339) (384,1.6441) (768,1.6626) (1536,1.4580) (3072,1.4351) (6144,1.4618) (12288,1.5550)};
        \addlegendentry{BERT-L12-H768-A12 (109.5M)}

    \nextgroupplot[title={\footnotesize BERT Family (Legal QA)}]
        \addplot[semithick, Set2bertL2H128A2, mark=*, mark size=1pt] coordinates {(32,5.3206) (64,3.9041) (128,3.3931) (256,3.3513) (512,3.3464) (1024,3.3935) (2048,3.4950)};
        \addplot[semithick, Set2bertL12H128A2, mark=*, mark size=1pt] coordinates {(32,3.8325) (64,2.8325) (128,2.5751) (256,2.4608) (512,2.4898) (1024,2.8382) (2048,3.5127)};
        \addplot[semithick, Set2bertL4H256A4, mark=*, mark size=1pt] coordinates {(32,4.0060) (64,2.5854) (128,2.2159) (256,1.9616) (512,1.7778) (1024,1.8747) (2048,1.9808) (4096,1.9266)};
        \addplot[semithick, Set2bertL4H512A8, mark=*, mark size=1pt] coordinates {(32,3.5204) (64,2.3355) (128,1.8768) (256,1.6756) (512,1.5272) (1024,1.5914) (2048,1.5533) (4096,1.5659) (8192,1.5958)};
        \addplot[semithick, Set2bertL2H768A12, mark=*, mark size=1pt] coordinates {(48,2.5404) (96,1.9164) (192,1.7426) (384,1.6601) (768,1.5194) (1536,1.5220) (3072,1.4802) (6144,1.4660) (12288,1.3696)};
        \addplot[semithick, Set2bertL8H512A8, mark=*, mark size=1pt] coordinates {(32,3.1642) (64,1.7863) (128,1.4604) (256,1.3201) (512,1.1876) (1024,1.1117) (2048,1.0939) (4096,1.1267) (8192,1.1408)};
        \addplot[semithick, Set2bertL12H768A12, mark=*, mark size=1pt] coordinates {(48,2.1675) (96,1.4360) (192,1.2164) (384,1.1193) (768,0.9307) (1536,1.0416) (3072,1.0028) (6144,1.0142) (12288,1.0831)};

    \nextgroupplot[
        title={\footnotesize Ettin Family (Scientific Papers)},
        legend to name=legendEttin,
    ]
        \addplot[semithick, Set2ettinL7H256A4, mark=*, mark size=1pt] coordinates {(64,2.3119) (128,2.1377) (256,2.0656) (512,2.0112) (1024,1.9447) (2048,1.9463) (4096,1.9507)};
        \addlegendentry{Ettin-L7-H256-A4 (16.8M)}
        \addplot[semithick, Set2ettinL10H384A6, mark=*, mark size=1pt] coordinates {(96,2.3094) (192,2.0620) (384,1.9575) (768,1.9344) (1536,1.9294) (3072,1.9108) (6144,1.8609)};
        \addlegendentry{Ettin-L10-H384-A6 (31.9M)}
        \addplot[semithick, Set2ettinL19H512A8, mark=*, mark size=1pt] coordinates {(128,2.2929) (256,2.1795) (512,2.1626) (1024,2.0946) (2048,2.0107) (4096,1.9533) (8192,2.0240)};
        \addlegendentry{Ettin-L19-H512-A8 (68.1M)}
        \addplot[semithick, Set2ettinL22H768A12, mark=*, mark size=1pt] coordinates {(192,2.0662) (384,1.7163) (768,1.6459) (1536,1.5743) (3072,1.5320) (6144,1.5486) (12288,1.5256)};
        \addlegendentry{Ettin-L22-H768-A12 (149.0M)}
        \addplot[semithick, Set2ettinL28H1024A16, mark=*, mark size=1pt] coordinates {(256,1.5705) (512,1.4234) (1024,1.1931) (2048,1.2818) (4096,1.2211) (8192,1.3177) (16384,1.2889)};
        \addlegendentry{Ettin-L28-H1024-A16 (394.8M)}
        \addplot[semithick, Set2ettinL28H1792A28, mark=*, mark size=1pt] coordinates {(448,1.3762) (896,1.2509) (1792,1.2208) (3584,1.1731) (7168,1.1607) (14336,1.1855) (28672,1.1058)};
        \addlegendentry{Ettin-L28-H1792-A28 (1028.1M)}

    \nextgroupplot[title={\footnotesize Ettin Family (Legal QA)}]
        \addplot[semithick, Set2ettinL7H256A4, mark=*, mark size=1pt] coordinates {(64,2.8820) (128,2.8804) (256,2.8720) (512,2.7471) (1024,2.6779) (2048,2.7241) (4096,2.6229)};
        \addplot[semithick, Set2ettinL10H384A6, mark=*, mark size=1pt] coordinates {(96,2.7676) (192,2.5538) (384,2.4681) (768,2.4685) (1536,2.4657) (3072,2.4666) (6144,2.3930)};
        \addplot[semithick, Set2ettinL19H512A8, mark=*, mark size=1pt] coordinates {(128,2.3279) (256,2.3913) (512,2.0711) (1024,2.1541) (2048,2.1341) (4096,2.2067) (8192,2.1275)};
        \addplot[semithick, Set2ettinL22H768A12, mark=*, mark size=1pt] coordinates {(192,2.0221) (384,2.0441) (768,1.7341) (1536,1.9848) (3072,1.9506) (6144,2.0087) (12288,1.9276)};
        \addplot[semithick, Set2ettinL28H1024A16, mark=*, mark size=1pt] coordinates {(256,1.8346) (512,1.6391) (1024,1.5668) (2048,1.6812) (4096,1.6706) (8192,1.6061) (16384,1.6121)};
        \addplot[semithick, Set2ettinL28H1792A28, mark=*, mark size=1pt] coordinates {(448,1.4928) (896,1.5913) (1792,1.4101) (3584,1.3004) (7168,1.4813) (14336,1.2454) (28672,1.2885)};
\end{groupplot}

\node[anchor=west, xshift=5pt] at (group c2r1.east) {\ref{legendBERT}};
\node[anchor=west, xshift=5pt] at (group c2r2.east) {\ref{legendEttin}};

\end{tikzpicture}
\caption{Out-of-domain evaluation on Paper Retrieval and Legal QA for BERT and Ettin families.}
\label{fig:ood_unified_figure}
\end{figure*}
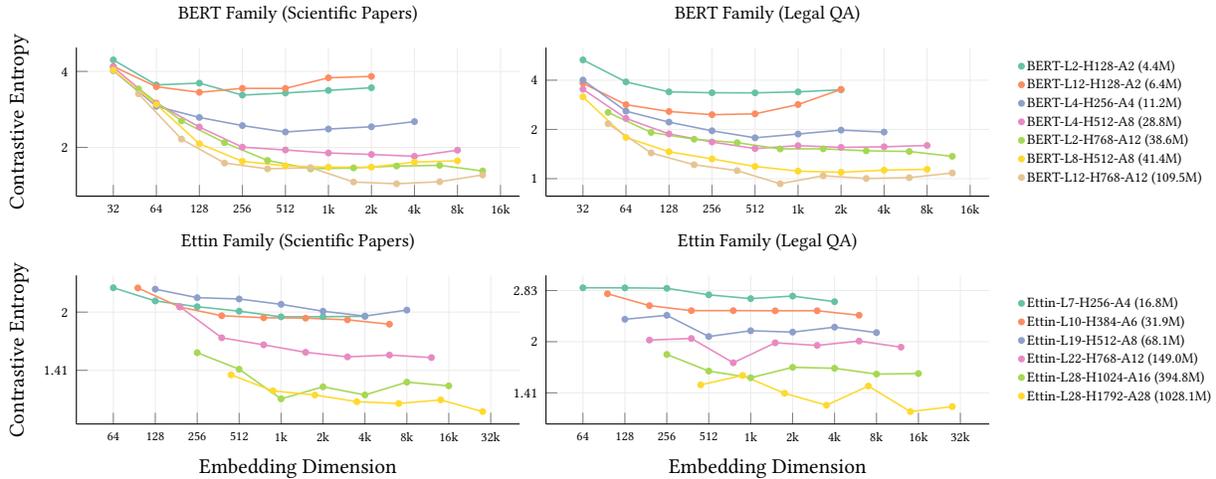

\subsection{Unified Model Parameter and Embedding Dimension Scaling Behavior}
The prior scaling law assumes that we have a fixed model, but as shown by prior work on scaling behavior of dense retrieval models \cite{scaling_laws_dense_retrieval}, the model size has a substantial impact on retrieval performance. With this in mind, we want to understand the interplay between these two parameters, which is why we use a range of model sizes within each model family while conducting our experiments. The results on the combined TREC DL and MSMARCO Dev evaluation sets for the BERT and Ettin model families can be seen in Figures \ref{fig:bert_joint_scaling_plot}
and \ref{fig:ettin_joint_scaling_plot}, respectively. The points represent the empirical results across model sizes as the embedding dimension is scaled. We observe that for similar embedding dimensions different model sizes have a significant impact on performance, although like scaling the embedding dimension the gains are diminishing. In this regard, our findings align with prior work on scaling dense retrieval model sizes while keeping the embedding dimension fixed \cite{scaling_laws_dense_retrieval}, which found the increase in performance from scaling model size follows a power law. 

An interesting result of the joint embedding dimension and model size plot is that we see across model sizes the scaling pattern is pretty similar. For example, across all model sizes in Figure \ref{fig:bert_joint_scaling_plot} we find that there is a sharp decrease from dimension 32 to 256 while beyond 1024 there is only a minimal decrease. This indicates that the required embedding dimension to do a task well, for a certain model scale, is similar and may be more related to the evaluation task than the model's hidden size. This interpretation aligns with the prior discussion about why there is only limited improvement as embedding dimension increases. Specifically, for a given task there is some amount of information that is required to do well, but at a certain point the common features are accounted for and additional embedding dimensions only helps for infrequent features which, by definition, do not appear in many evaluation tasks.

Another takeaway is that it is very difficult to make up for limited model size with embedding dimension and vice versa. Once the embedding dimension is beyond a critical point where improvement begins to plateau, trying to match a larger model's performance by scaling the embedding dimension becomes futile. For example in the TREC DL portion of Figure \ref{fig:bert_joint_scaling_plot}, the BERT-L4-H512 model can scale to an embedding dimension of 8k, but is still worse than the larger BERT-L8-H512 with an embedding dimension of 256. The same is also true for embedding dimension, but mostly at smaller embedding dimensions. For instance, when BERT-L8-H512 has an embedding dimension of 32 even the smallest BERT model can get lower performance with a larger embedding dimension, while this is not true when BERT-L8-H512 has any larger embedding dimension.

Given the empirical results, as in the dimension only case, it is useful to have a predictable equation for contrastive entropy given both model size and embedding dimension. We do so by adding another term to the existing equation which accounts for the improved performance as model size increases:
\setlength{\abovedisplayskip}{3pt}
\setlength{\belowdisplayskip}{3pt}
\begin{equation} \label{eq:joint_scaling_laws}
    L(D, N) = \frac{A}{D^{\alpha}} + \frac{B}{N^{\beta}} + \delta
\end{equation}
like Equation \eqref{eq:dim_only_scaling_law} $D$ represents the embedding dimension, while $N$ represents the model size based on the total number of parameters. The values of $A$, $B$, $\alpha$, $\beta$, and $\delta$ are learned parameters. As before, $\delta$ represents the irreducible error of the task. We use the same approach to fit this scaling law as the dimension only scaling law. When fitting the scaling law we divide the model parameters by one million (e.g. 1,000,000 become 1.0) to reduce the scale of the parameters.

Applying the joint scaling law to our empirical results for the BERT model family on the combined TREC DL dataset and MSMARCO Dev can be seen in Figure \ref{fig:bert_joint_scaling_plot} with the dashed lines. Similarly, the results for the Ettin model family are shown in Figure \ref{fig:ettin_joint_scaling_plot}.

One useful insight from our joint scaling laws is that \textbf{in many cases it is better to use a larger model with a smaller embedding dimension than a smaller model with a larger embedding dimension.} Although the exact trade-off is task specific, this finding is repeated across the models and evaluation tasks. A meaningful application of this is that embedding dimension is often the largest controllable factor in the size of the retrieval index and using smaller embeddings with a larger model could keep the same quality while halving the index size. We formalize a related idea in Section \ref{sec:cost_aware_analysis} where we find the optimal model size and embedding dimension for a fixed inference FLOPs budget.

\subsection{Scaling Behavior for Aligned and Unaligned Evaluation Tasks}

In the results shown so far, we focus on the contrastive entropy of ``aligned'' evaluation tasks. These are tasks that align with how the model was trained, including the training data and training objective. This practice is in line with prior work \cite{scaling_laws_dense_retrieval, scaling_laws_neural_language_models, chinchilla_compute_optimal_llms}, but understanding how scaling impacts out-of-domain evaluation tasks is of significant importance for building generalizable retrieval systems. In this section, we explore the scaling behaviors on out-of-domain evaluation tasks. 

The results of evaluating both the Ettin and BERT model families on our two out-of-domain datasets Legal QA and Paper Retrieval are presented in Figure \ref{fig:ood_unified_figure}. The results show that generally as embedding dimension increases there is a decrease in contrastive entropy which matches the aligned results. However, unlike the aligned results the behavior is more erratic without an obvious predictable pattern. Another unique characteristic is several of the models begin to have contrastive entropy tick up as embedding dimension increases. This behavior is especially common for the BERT model on the Paper Retrieval task. One explanation is that the Paper Retrieval task is more out-of-distribution for the BERT models which are only trained on short natural language questions, but this would not explain the increase as embedding dimensions become larger. Another explanation that can better account for this uptick is that as the embedding dimension increases a more complex similarity function can be learned and as the BERT models are trained with knowledge distillation the larger embedding dimensions may become more aligned with the teacher and less able to generalize. This hypothesis is supported by the absence of an uptick in the Ettin models which do not use knowledge distillation and prior work by \citet{scaling_sparse_dense_decoder_llms} that shows knowledge distillation can hurt generalization.

Even with the upticks in the BERT models the general trend is still that performance improves as we scale dimensions. \textbf{This means the scaling laws likely still broadly apply out-of-domain, just  not as predictably.} These results mean scaling is likely a robust way to improve dense retrieval performance in a generalized way. 

\subsection{Scaling Behavior Based on Ranking Metrics}
\label{sec:ir_metric_scaling}
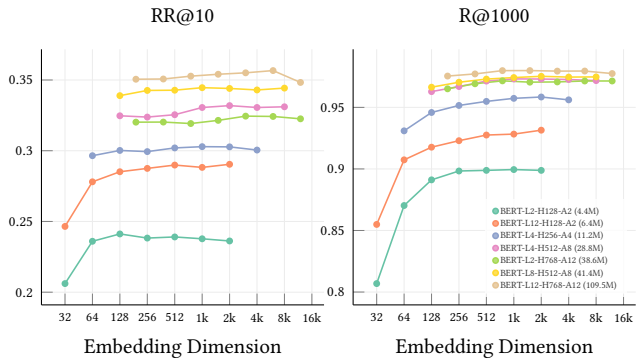
\begin{figure}[tp]
\centering
\definecolor{Set2bertL2H128A2}{rgb}{0.400, 0.761, 0.647}
\definecolor{Set2bertL12H128A2}{rgb}{0.988, 0.553, 0.384}
\definecolor{Set2bertL4H256A4}{rgb}{0.553, 0.627, 0.796}
\definecolor{Set2bertL4H512A8}{rgb}{0.906, 0.541, 0.765}
\definecolor{Set2bertL2H768A12}{rgb}{0.651, 0.847, 0.329}
\definecolor{Set2bertL8H512A8}{rgb}{1.000, 0.851, 0.184}
\definecolor{Set2bertL12H768A12}{rgb}{0.898, 0.769, 0.580}
\begin{tikzpicture}
\begin{groupplot}[
    group style={
        group size=2 by 1,
        horizontal sep=11pt,
        xlabels at=edge bottom,
        ylabels at=edge left,
    },
    width=0.30\textwidth, height=0.28\textwidth,
    xmode=log, %
    log basis x={2}, %
    tick label style={font=\tiny},
    xlabel={\small Embedding Dimension},
    grid=both, grid style={line width=.1pt, draw=gray!15},
    axis lines*=left,
    xtick={32,64,128,256,512,1024,2048,4096,8192,16384,32768},
    xticklabels={32,64,128,256,512,1k,2k,4k,8k,16k,32k},
    log ticks with fixed point,
    yticklabel style={/pgf/number format/fixed, /pgf/number format/precision=2},
]
    \nextgroupplot[
        title={\small RR@10},
        legend style={at={(0.97,0.03)}, anchor=south east, nodes={scale=0.7, transform shape}, fill opacity=0.8, draw=none, legend columns=1, legend image code/.code={\draw[#1] plot coordinates {(0cm,0cm)};}},
        legend cell align={left},
    ]
        \addplot[semithick, Set2bertL2H128A2, mark=*, mark size=1pt] coordinates {(32, 0.2060842770728154) (64, 0.23598853868194755) (128, 0.24117558102515044) (256, 0.23826050620821285) (512, 0.23904545868012825) (1024, 0.23775213762677863) (2048, 0.23620792968572227)};
        \addplot[semithick, Set2bertL12H128A2, mark=*, mark size=1pt] coordinates {(32, 0.24648377450311476) (64, 0.2780221835630131) (128, 0.2851422999954506) (256, 0.2875076181379896) (512, 0.2898607131486775) (1024, 0.2882200845954413) (2048, 0.2904076272342736)};
        \addplot[semithick, Set2bertL4H256A4, mark=*, mark size=1pt] coordinates {(64, 0.29648678764724473) (128, 0.30019858325373944) (256, 0.29937991767862726) (512, 0.3019783849547452) (1024, 0.30287226542957063) (2048, 0.3027251898849304) (4096, 0.3004764747350691)};
        \addplot[semithick, Set2bertL4H512A8, mark=*, mark size=1pt] coordinates {(128, 0.3246962409605665) (256, 0.3237861008777901) (512, 0.3254599877200159) (1024, 0.33055669259107595) (2048, 0.3318823054532213) (4096, 0.3305635716559781) (8192, 0.33105107563560165)};
        \addplot[semithick, Set2bertL2H768A12, mark=*, mark size=1pt] coordinates {(192, 0.32018965752489975) (384, 0.32025395688361113) (768, 0.31921851548642177) (1536, 0.3214609200891418) (3072, 0.3244477986992301) (6144, 0.3242295447309751) (12288, 0.322552019375083)};
        \addplot[semithick, Set2bertL8H512A8, mark=*, mark size=1pt] coordinates {(128, 0.3389324942011186) (256, 0.34261586391958787) (512, 0.34278056351480346) (1024, 0.3445678696502477) (2048, 0.34402175148951647) (4096, 0.3429575430936453) (8192, 0.3442642811661429)};
        \addplot[semithick, Set2bertL12H768A12, mark=*, mark size=1pt] coordinates {(192, 0.35054048983490227) (384, 0.3506975712921273) (768, 0.3527197889662071) (1536, 0.35405597625869767) (3072, 0.35506498158002453) (6144, 0.3566530222404147) (12288, 0.34827261563651274)};

    \nextgroupplot[
        title={\small R@1000},
        legend style={at={(0.97,0.03)}, anchor=south east, nodes={scale=0.4, transform shape}, fill opacity=0.8, draw=none, legend columns=1, legend image code/.code={\draw[#1] plot coordinates {(0cm,0cm)};}},
        legend cell align={left},
    ]
        \addplot[semithick, Set2bertL2H128A2, mark=*, mark size=1pt] coordinates {(32,0.8068648519579752) (64,0.8702722063037249) (128,0.8911294173829992) (256, 0.8983285577841452) (512, 0.8988419293218722) (1024, 0.8994866284622733) (2048, 0.8988299904489017)};
        \addlegendentry{BERT-L2-H128-A2 (4.4M)}
        \addplot[semithick, Set2bertL12H128A2, mark=*, mark size=1pt] coordinates {(32,0.8549188156638015) (64,0.907378223495702) (128,0.9176695319961795) (256,0.9229345749761222) (512,0.9275310410697231) (1024,0.9282831900668578) (2048, 0.9314828080229226)};
        \addlegendentry{BERT-L12-H128-A2 (6.4M)}
        \addplot[semithick, Set2bertL4H256A4, mark=*, mark size=1pt] coordinates {(64,0.9309455587392551) (128,0.9458572110792742) (256,0.9515759312320918) (512,0.9548591212989495) (1024,0.957342406876791) (2048,0.9584288443170965) (4096, 0.9561127029608404)};
        \addlegendentry{BERT-L4-H256-A4 (11.2M)}
        \addplot[semithick, Set2bertL4H512A8, mark=*, mark size=1pt] coordinates {(128,0.9628104106972304) (256,0.9669412607449857) (512,0.971370582617001) (1024,0.9731375358166189) (2048,0.9730300859598854) (4096,0.9726361031518626) (8192,0.9717526265520535)};
        \addlegendentry{BERT-L4-H512-A8 (28.8M)}
        \addplot[semithick, Set2bertL2H768A12, mark=*, mark size=1pt] coordinates {(192,0.9650787965616046) (384,0.9691857688634192) (768,0.9715496657115568) (1536,0.970487106017192) (3072,0.9706781279847183) (6144,0.9714422158548233) (12288,0.9714422158548235)};
        \addlegendentry{BERT-L2-H768-A12 (38.6M)}
        \addplot[semithick, Set2bertL8H512A8, mark=*, mark size=1pt] coordinates {(128,0.966404011461318) (256,0.9704274116523401) (512,0.972982330468004) (1024,0.9742239732569247) (2048,0.975238777459408) (4096,0.9746179560649476) (8192,0.9747492836676219)};
        \addlegendentry{BERT-L8-H512-A8 (41.4M)}
        \addplot[semithick, Set2bertL12H768A12, mark=*, mark size=1pt] coordinates {(192,0.9754417382999045) (384,0.9771728748806112) (768,0.9798591212989495) (1536,0.9798949379178604) (3072,0.9794651384909264) (6144,0.9795248328557785) (12288,0.9774832855778416)};
        \addlegendentry{BERT-L12-H768-A12 (109.5M)}
    
\end{groupplot}
\end{tikzpicture}
\caption{Scaling behavior based on ranking metrics (RR@10 and R@1000) for BERT model family on MSMARCO Dev.}
\label{fig:ir_metrics_figure}
\vspace{-5pt}
\end{figure}

While contrastive entropy serves as our primary evaluation metric, it is important to assess whether the observed scaling trends translate to improvements in a full retrieval setting. To verify that this correlation holds in our experimental setup, in this section we run an auxiliary retrieval evaluation and analyze scaling behavior based on classic retrieval metrics. Since we use contrastive entropy as a proxy for retrieval performance and exhaustive retrieval evaluation across all models and datasets would fall outside the scope of our study, we restrict this analysis to a representative subset of our experimental configurations. Specifically, we evaluate the BERT model family on the MSMARCO Dev dataset and report retrieval effectiveness in terms of reciprocal rank (RR@10) and recall (R@1000) as the embedding dimension is scaled. The results of this evaluation are presented in Figure \ref{fig:ir_metrics_figure}. We can observe that the trend of performance improvement holds as we scale the embedding dimension, though the scaling behavior follows an increasing trend since we switched from contrastive entropy to ranking metrics. As expected from using discrete measures, the scaling behavior is not as smooth as depicted by contrastive entropy. This can be explained by the difference between the measurement objective of contrastive entropy and metrics such as R@1000 and RR@10. While ranking metrics evaluate the order of retrieved documents, they do not consider performance improvement beyond their cutoff point (10 for reciprocal rank and 1000 for recall in our setting), and they do not pick up on subtle improvements on document scores that do not alter the ranking order. 
Interestingly, this effect is especially evident when comparing the results in terms of RR@10 and R@1000. Reciprocal rank is inherently more restrictive in measuring retrieval performance here as it depends solely on the rank of the first relevant document and is capped at a cutoff of 10. In contrast, R@1000 accounts for all relevant documents retrieved within a substantially larger cutoff, allowing it to capture performance gains across a wider portion of the ranking. Consequently, the scaling trend with embedding dimension is more clearly reflected in R@1000 than in RR@10.

\subsection{Cost Aware Embedding Dimension Selection} \label{sec:cost_aware_analysis}
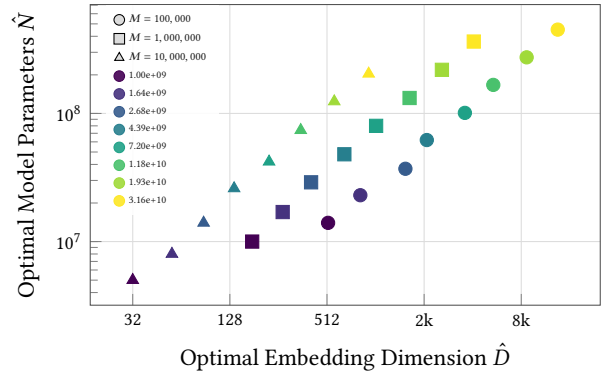
\begin{figure}[tp]
  \centering
  \definecolor{SoftGridColor}{RGB}{220, 220, 220}
  \definecolor{AxisFrameColor}{RGB}{80, 80, 80}
  \definecolor{color_1000000000}{HTML}{440154}
  \definecolor{color_1637893706}{HTML}{46327E}
  \definecolor{color_2682695795}{HTML}{365C8D}
  \definecolor{color_4393970560}{HTML}{277F8E}
  \definecolor{color_7196856730}{HTML}{1FA187}
  \definecolor{color_11787686347}{HTML}{4AC16D}
  \definecolor{color_19306977288}{HTML}{A0DA39}
  \definecolor{color_31622776601}{HTML}{FDE725}
  \begin{tikzpicture}
    \begin{axis}[
      xmode=log, log basis x={2}, 
      ymode=log,
      scale only axis, grid=major,
      grid style={solid, SoftGridColor},
      width=0.8\columnwidth, height=4cm,
      tick pos=left, tick label style={font=\footnotesize},
      xtick={32,128,512,2048,8192},
      xticklabels={32,128,512,2k,8k},
      legend style={at={(0.03,1)}, anchor=north west, font=\tiny, draw=SoftGridColor, cells={anchor=west}, nodes={scale=0.8, transform shape}, fill opacity=0.9, draw=none},
      axis line style={AxisFrameColor},
      xlabel={Optimal Embedding Dimension $\hat{D}$},
      ylabel={Optimal Model Parameters $\hat{N}$},
    ]
      \addlegendimage{only marks, mark=*, draw=black, fill=gray!30}
      \addlegendentry{$M=100,000$}
      \addlegendimage{only marks, mark=square*, draw=black, fill=gray!30}
      \addlegendentry{$M=1,000,000$}
      \addlegendimage{only marks, mark=triangle*, draw=black, fill=gray!30}
      \addlegendentry{$M=10,000,000$}
      \addlegendimage{only marks, mark=*, color_1000000000}
      \addlegendentry{1.00e+09}
      \addlegendimage{only marks, mark=*, color_1637893706}
      \addlegendentry{1.64e+09}
      \addlegendimage{only marks, mark=*, color_2682695795}
      \addlegendentry{2.68e+09}
      \addlegendimage{only marks, mark=*, color_4393970560}
      \addlegendentry{4.39e+09}
      \addlegendimage{only marks, mark=*, color_7196856730}
      \addlegendentry{7.20e+09}
      \addlegendimage{only marks, mark=*, color_11787686347}
      \addlegendentry{1.18e+10}
      \addlegendimage{only marks, mark=*, color_19306977288}
      \addlegendentry{1.93e+10}
      \addlegendimage{only marks, mark=*, color_31622776601}
      \addlegendentry{3.16e+10}
      \addplot[only marks, forget plot, mark=*, color=color_1000000000, mark options={fill=color_1000000000, scale=1.25}] coordinates {(520.0,14000000.0)};
      \addplot[only marks, forget plot, mark=*, color=color_1637893706, mark options={fill=color_1637893706, scale=1.25}] coordinates {(824.0,23000000.0)};
      \addplot[only marks, forget plot, mark=*, color=color_2682695795, mark options={fill=color_2682695795, scale=1.25}] coordinates {(1568.0,37000000.0)};
      \addplot[only marks, forget plot, mark=*, color=color_4393970560, mark options={fill=color_4393970560, scale=1.25}] coordinates {(2128.0,62000000.0)};
      \addplot[only marks, forget plot, mark=*, color=color_7196856730, mark options={fill=color_7196856730, scale=1.25}] coordinates {(3664.0,101000000.0)};
      \addplot[only marks, forget plot, mark=*, color=color_11787686347, mark options={fill=color_11787686347, scale=1.25}] coordinates {(5496.0,167000000.0)};
      \addplot[only marks, forget plot, mark=*, color=color_19306977288, mark options={fill=color_19306977288, scale=1.25}] coordinates {(8848.0,274000000.0)};
      \addplot[only marks, forget plot, mark=*, color=color_31622776601, mark options={fill=color_31622776601, scale=1.25}] coordinates {(13792.0,451000000.0)};
      \addplot[only marks, forget plot, mark=square*, color=color_1000000000, mark options={fill=color_1000000000, scale=1.25}] coordinates {(176.0,10000000.0)};
      \addplot[only marks, forget plot, mark=square*, color=color_1637893706, mark options={fill=color_1637893706, scale=1.25}] coordinates {(272.0,17000000.0)};
      \addplot[only marks, forget plot, mark=square*, color=color_2682695795, mark options={fill=color_2682695795, scale=1.25}] coordinates {(408.0,29000000.0)};
      \addplot[only marks, forget plot, mark=square*, color=color_4393970560, mark options={fill=color_4393970560, scale=1.25}] coordinates {(656.0,48000000.0)};
      \addplot[only marks, forget plot, mark=square*, color=color_7196856730, mark options={fill=color_7196856730, scale=1.25}] coordinates {(1032.0,80000000.0)};
      \addplot[only marks, forget plot, mark=square*, color=color_11787686347, mark options={fill=color_11787686347, scale=1.25}] coordinates {(1664.0,132000000.0)};
      \addplot[only marks, forget plot, mark=square*, color=color_19306977288, mark options={fill=color_19306977288, scale=1.25}] coordinates {(2640.0,219000000.0)};
      \addplot[only marks, forget plot, mark=square*, color=color_31622776601, mark options={fill=color_31622776601, scale=1.25}] coordinates {(4160.0,364000000.0)};
      \addplot[only marks, forget plot, mark=triangle*, color=color_1000000000, mark options={fill=color_1000000000, scale=1.25}] coordinates {(32.0,5000000.0)};
      \addplot[only marks, forget plot, mark=triangle*, color=color_1637893706, mark options={fill=color_1637893706, scale=1.25}] coordinates {(56.0,8000000.0)};
      \addplot[only marks, forget plot, mark=triangle*, color=color_2682695795, mark options={fill=color_2682695795, scale=1.25}] coordinates {(88.0,14000000.0)};
      \addplot[only marks, forget plot, mark=triangle*, color=color_4393970560, mark options={fill=color_4393970560, scale=1.25}] coordinates {(136.0,26000000.0)};
      \addplot[only marks, forget plot, mark=triangle*, color=color_7196856730, mark options={fill=color_7196856730, scale=1.25}] coordinates {(224.0,42000000.0)};
      \addplot[only marks, forget plot, mark=triangle*, color=color_11787686347, mark options={fill=color_11787686347, scale=1.25}] coordinates {(352.0,74000000.0)};
      \addplot[only marks, forget plot, mark=triangle*, color=color_19306977288, mark options={fill=color_19306977288, scale=1.25}] coordinates {(568.0,124000000.0)};
      \addplot[only marks, forget plot, mark=triangle*, color=color_31622776601, mark options={fill=color_31622776601, scale=1.25}] coordinates {(928.0,204000000.0)};
    \end{axis}
  \end{tikzpicture}
  \caption{Optimal $(\hat{N}, \hat{D})$ under various FLOPs budgets and corpus sizes $M$. Across corpus sizes as FLOPs budget increases, both embedding size and model size increase, suggesting both are required for optimal performance.}
  \label{fig:optimal_params}
  \vspace{-5pt}
\end{figure}

A useful application of our joint scaling law is that we can find the optimal trade-off between model size and embedding dimension at various compute budgets. In this section, we study the end-to-end trade-off between model capacity and retrieval cost under a fixed per-query FLOPs budget. Concretely, for each query we (i) run the encoder to produce a query embedding and (ii) score that embedding against a corpus of $M$ precomputed document vectors. Let $B$ denote the total FLOPs budget per query, $T$ the query sequence length (tokens), $N$ the number of model parameters, and $D$ the embedding dimension. Following standard inference-cost approximations, we model the query-encoding cost as $C_{\text{enc}}(N,T) \approx 2NT$ FLOPs, and scoring as $C_{\text{score}}(M,D)$ FLOPs (counting one multiply-add per dimension per scored document). We enforce the compute budget $C_{\text{enc}} + C_{\text{score}} \le B$ by sweeping an allocation parameter $\gamma \in (0,1)$ that splits compute between encoding and scoring: $C_{\text{enc}}=\gamma B$ and $C_{\text{score}}=(1-\gamma)B$, and then selecting the configuration that minimizes the predicted contrastive entropy under our fitted joint scaling law $\widehat{L}(D,N)$:
\[
\begin{gathered}
\hat{\gamma} \in \arg\min_{\gamma \in (0,1)} \widehat{L}(D(\gamma),N(\gamma)),\quad
(\hat{N}, \hat{D}) = (N(\hat{\gamma}), D(\hat{\gamma})).
\end{gathered}
\]
\paragraph{Exhaustive scoring.}
For exhaustive (brute force) scoring over the full corpus, we use $C_{\text{score}}(M,D) \approx 2MD$. Under the $\gamma$-split, this yields
\[
N(\gamma)=\frac{\gamma B}{2T}
\qquad\text{and}\qquad
D(\gamma)=\frac{(1-\gamma)B}{2M}.
\]

\paragraph{ANN-style scoring.}
If approximate nearest neighbor (ANN) search is used, $M$ can be replaced by an index-dependent proxy for the expected number of inner products evaluated per query. As a simple illustrative approximation, we can model the scoring cost as
\[
C_{\text{score}}(M,D) \approx 2D\log(M),
\]
which corresponds to evaluating a logarithmic number of candidate comparisons in $M$ (up to method- and implementation-specific constants). Under the $\gamma$-split, this gives
\[
N(\gamma)=\frac{\gamma B}{2T}
\qquad\text{and}\qquad
D(\gamma)=\frac{(1-\gamma)B}{2\log(M)}.
\]

\input{plot_sources/flops_budget_embedding_versus_entropy}

To keep this analysis tractable, we make some simplifying assumptions. First, we do not account for the projection to the final embedding dimension $D$ in the encoding FLOPs $C_{enc}$. With that said, for most optimal setups the embedding dimension projection would be a very minor part of the overall encoding. We also assume ANN scoring would produce similar retrieval results to the exhaustive scoring. Although this is generally never completely true, in many use cases approximate approaches can be highly competitive with exhaustive approaches \cite{anns_comprehensive_survey}.

The results of the optimal embedding size and model size, with $M = \{10^{5}, 10^{6}, 10^{7}\}$ and  $T=32$, can be seen in Figure \ref{fig:optimal_params}. We use the joint scaling law parameters fit on BERT using TREC DL Combined. The results show that across corpus sizes there is a clear pattern that for optimal FLOP allocation both model size and embedding dimension should be increased jointly. Intuitively, this make sense as both axes, embedding dimension and model size, have diminishing returns as they are scaled so it is sensible the optimal approach scales both proportionally. We also see that for larger collections (i.e. larger values of $M$) the optimal embedding is smaller to account for the additional cost of scoring more documents.

Figure \ref{fig:flops_embedding_dim_vs_contrastive_entropy} shows the impact of varying the embedding dimension on the predicted contrastive entropy for different inference FLOP budgets. We use the joint scaling law parameters fit on BERT using TREC DL Combined. \textbf{We observe that at low embedding dimensions, contrastive entropy is high because the embedding dimension is a bottleneck. As embedding dimension decreases so does the contrastive entropy until an inflection point where the contrastive entropy rapidly increases. This inflection point represent the optimal embedding dimension for a given FLOP budget.} After that point the scoring becomes too expensive due to the embedding size and takes away model size which results in the sudden increase in contrastive entropy. The dashed lines represent the same setup as the solid lines but with approximate nearest neighbor scoring which we assume requires only $log(M)$ inner-products. In the approximate setup, we can see that the optimal embedding dimension is far higher than with exhaustive retrieval, in fact the inflection point is not even visible on the graph. This means that with ANN scoring it is possible to substantially improve performance with the same FLOPs budget, this is especially when using smaller budgets. \textbf{This insight might make a more efficient, but less effective, ANN approach paired with a larger embedding dimension preferable to the more conventional approach of using a embedding dimension inherited from the hidden state size with a high quality ANN approach.} 

Our preliminary analysis is only the first step in applying the joint scaling laws we derived to inform dense retrieval modeling choices. Even with just this preliminary analysis, we have found interesting patterns and design guidelines for future developers of dense retrieval models. We believe that our scaling laws will be instrumental in scaling dense retrieval for the demanding new tasks of the future.

\section{Conclusion}
In this work, we presented a comprehensive empirical analysis of the relationship between embedding dimension and dense retrieval performance. By treating embedding dimension as an independent scaling axis rather than a static byproduct of the encoder architecture, we demonstrated that retrieval effectiveness follows a predictable power law as dimensionality increases. We validated this finding across two distinct model families, BERT and Ettin, and derived a joint scaling law that unifies model size and embedding dimension into a single predictive framework.

Our results offer critical insights for the design of efficient retrieval systems. We found that while increasing embedding dimension generally improves performance, it does so with diminishing returns. Crucially, our cost-aware analysis reveals that the standard practice of using the encoder's native hidden size is rarely optimal. In many resource-constrained scenarios, utilizing a larger model with a projected, smaller embedding dimension yields superior retrieval quality compared to a smaller model with a larger embedding. Furthermore, we showed that when coupled with approximate nearest neighbor search, scaling embedding dimensions beyond the native hidden size can offer significant performance gains for a fixed compute budget.

While our scaling laws hold strongly for tasks aligned with the training distribution, our evaluation on out-of-domain collections suggests that scaling behavior becomes less predictable when generalization is required, particularly for models trained with knowledge distillation. This highlights the importance of aligning model capacity with the complexity of the target retrieval task.

There are many avenues for future work in this area. A natural extension of this study is to investigate scaling laws for sparse embeddings, specifically analyzing the interplay between high-dimensional sparse representations and their sparsity ratios. Additionally, further research is needed to develop training objectives that encourage models to more uniformly utilize the geometric capacity of their embedding space, ensuring that every added dimension contributes meaningfully to retrieval effectiveness. We hope that the scaling laws and design guidelines presented in this work serve as a foundation for building the next generation of effective and efficient neural retrieval systems.

\begin{acks}
This work was supported in part by the Center for Intelligent Information Retrieval, in part by NSF grant \#2402873, in part by the Office of Naval Research contract \#N000142412612, and in part by the NSF Graduate Research Fellowships Program (GRFP) Award \#1938059. Any opinions, findings and conclusions or recommendations expressed in this material are those of the authors and do not necessarily reflect those of the sponsor.
\end{acks}

\bibliographystyle{ACM-Reference-Format}
\bibliography{main}


\begin{thebibliography}{39}


\ifx \showCODEN    \undefined \def \showCODEN     #1{\unskip}     \fi
\ifx \showISBNx    \undefined \def \showISBNx     #1{\unskip}     \fi
\ifx \showISBNxiii \undefined \def \showISBNxiii  #1{\unskip}     \fi
\ifx \showISSN     \undefined \def \showISSN      #1{\unskip}     \fi
\ifx \showLCCN     \undefined \def \showLCCN      #1{\unskip}     \fi
\ifx \shownote     \undefined \def \shownote      #1{#1}          \fi
\ifx \showarticletitle \undefined \def \showarticletitle #1{#1}   \fi
\ifx \showURL      \undefined \def \showURL       {\relax}        \fi
\providecommand\bibfield[2]{#2}
\providecommand\bibinfo[2]{#2}
\providecommand\natexlab[1]{#1}
\providecommand\showeprint[2][]{arXiv:#2}

\bibitem[Craswell et~al\mbox{.}(2020a)]%
        {trec_2020_deep_learning}
\bibfield{author}{\bibinfo{person}{Nick Craswell}, \bibinfo{person}{Bhaskar Mitra}, \bibinfo{person}{Emine Yilmaz}, {and} \bibinfo{person}{Daniel Campos}.} \bibinfo{year}{2020}\natexlab{a}.
\newblock \showarticletitle{Overview of the {TREC} 2020 deep learning track}. In \bibinfo{booktitle}{\emph{Proceedings of the Twenty-Ninth Text REtrieval Conference (TREC 2020)}}. \bibinfo{publisher}{NIST}.
\newblock


\bibitem[Craswell et~al\mbox{.}(2020b)]%
        {trec_dl_2020}
\bibfield{author}{\bibinfo{person}{Nick Craswell}, \bibinfo{person}{Bhaskar Mitra}, \bibinfo{person}{Emine Yilmaz}, {and} \bibinfo{person}{Daniel Campos}.} \bibinfo{year}{2020}\natexlab{b}.
\newblock \showarticletitle{Overview of the {TREC} 2020 Deep Learning Track}. In \bibinfo{booktitle}{\emph{Proceedings of the Twenty-Ninth Text REtrieval Conference, {TREC} 2020, Virtual Event [Gaithersburg, Maryland, USA], November 16-20, 2020}} \emph{(\bibinfo{series}{{NIST} Special Publication}, Vol.~\bibinfo{volume}{1266})}, \bibfield{editor}{\bibinfo{person}{Ellen~M. Voorhees} {and} \bibinfo{person}{Angela Ellis}} (Eds.). \bibinfo{publisher}{National Institute of Standards and Technology {(NIST)}}.
\newblock
\urldef\tempurl%
\url{https://trec.nist.gov/pubs/trec29/papers/OVERVIEW.DL.pdf}
\showURL{%
\tempurl}


\bibitem[Craswell et~al\mbox{.}(2021)]%
        {trec_dl_21}
\bibfield{author}{\bibinfo{person}{Nick Craswell}, \bibinfo{person}{Bhaskar Mitra}, \bibinfo{person}{Emine Yilmaz}, \bibinfo{person}{Daniel Campos}, {and} \bibinfo{person}{Jimmy Lin}.} \bibinfo{year}{2021}\natexlab{}.
\newblock \showarticletitle{Overview of the {TREC} 2021 Deep Learning Track}. In \bibinfo{booktitle}{\emph{Proceedings of the Thirtieth Text REtrieval Conference, {TREC} 2021, online, November 15-19, 2021}} \emph{(\bibinfo{series}{{NIST} Special Publication}, Vol.~\bibinfo{volume}{500-335})}, \bibfield{editor}{\bibinfo{person}{Ian Soboroff} {and} \bibinfo{person}{Angela Ellis}} (Eds.). \bibinfo{publisher}{National Institute of Standards and Technology {(NIST)}}.
\newblock
\urldef\tempurl%
\url{https://trec.nist.gov/pubs/trec30/papers/Overview-DL.pdf}
\showURL{%
\tempurl}


\bibitem[Devlin et~al\mbox{.}(2019)]%
        {bert_bidirectional_transformers}
\bibfield{author}{\bibinfo{person}{Jacob Devlin}, \bibinfo{person}{Ming-Wei Chang}, \bibinfo{person}{Kenton Lee}, {and} \bibinfo{person}{Kristina Toutanova}.} \bibinfo{year}{2019}\natexlab{}.
\newblock \showarticletitle{{BERT}: Pre-training of Deep Bidirectional Transformers for Language Understanding}. In \bibinfo{booktitle}{\emph{Proceedings of the 2019 Conference of the North {A}merican Chapter of the Association for Computational Linguistics: Human Language Technologies, Volume 1 (Long and Short Papers)}}. \bibinfo{publisher}{Association for Computational Linguistics}, \bibinfo{pages}{4171--4186}.
\newblock
\href{https://doi.org/10.18653/v1/N19-1423}{doi:\nolinkurl{10.18653/v1/N19-1423}}


\bibitem[Douze et~al\mbox{.}(2024)]%
        {faiss_library}
\bibfield{author}{\bibinfo{person}{Matthijs Douze}, \bibinfo{person}{Alexandr Guzhva}, \bibinfo{person}{Chengqi Deng}, \bibinfo{person}{Jeff Johnson}, \bibinfo{person}{Gergely Szilvasy}, \bibinfo{person}{Pierre{-}Emmanuel Mazar{\'{e}}}, \bibinfo{person}{Maria Lomeli}, \bibinfo{person}{Lucas Hosseini}, {and} \bibinfo{person}{Herv{\'{e}} J{\'{e}}gou}.} \bibinfo{year}{2024}\natexlab{}.
\newblock \showarticletitle{The Faiss library}.
\newblock \bibinfo{journal}{\emph{CoRR}}  \bibinfo{volume}{abs/2401.08281} (\bibinfo{year}{2024}).
\newblock
\showeprint[arXiv]{2401.08281}
\href{https://doi.org/10.48550/ARXIV.2401.08281}{doi:\nolinkurl{10.48550/ARXIV.2401.08281}}


\bibitem[Fang et~al\mbox{.}(2024)]%
        {scaling_laws_dense_retrieval}
\bibfield{author}{\bibinfo{person}{Yan Fang}, \bibinfo{person}{Jingtao Zhan}, \bibinfo{person}{Qingyao Ai}, \bibinfo{person}{Jiaxin Mao}, \bibinfo{person}{Weihang Su}, \bibinfo{person}{Jia Chen}, {and} \bibinfo{person}{Yiqun Liu}.} \bibinfo{year}{2024}\natexlab{}.
\newblock \showarticletitle{Scaling Laws For Dense Retrieval}.
\newblock \bibinfo{journal}{\emph{CoRR}}  \bibinfo{volume}{abs/2403.18684} (\bibinfo{year}{2024}).
\newblock
\showeprint[arXiv]{2403.18684}
\href{https://doi.org/10.48550/ARXIV.2403.18684}{doi:\nolinkurl{10.48550/ARXIV.2403.18684}}


\bibitem[Gao and Callan(2021)]%
        {condenser_pretraining_dense_retrieval}
\bibfield{author}{\bibinfo{person}{Luyu Gao} {and} \bibinfo{person}{Jamie Callan}.} \bibinfo{year}{2021}\natexlab{}.
\newblock \showarticletitle{Condenser: a Pre-training Architecture for Dense Retrieval}. In \bibinfo{booktitle}{\emph{Proceedings of the 2021 Conference on Empirical Methods in Natural Language Processing}}. \bibinfo{publisher}{Association for Computational Linguistics}, \bibinfo{pages}{981--993}.
\newblock
\href{https://doi.org/10.18653/v1/2021.emnlp-main.75}{doi:\nolinkurl{10.18653/v1/2021.emnlp-main.75}}


\bibitem[Hoffmann et~al\mbox{.}(2022)]%
        {chinchilla_compute_optimal_llms}
\bibfield{author}{\bibinfo{person}{Jordan Hoffmann}, \bibinfo{person}{Sebastian Borgeaud}, \bibinfo{person}{Arthur Mensch}, \bibinfo{person}{Elena Buchatskaya}, \bibinfo{person}{Trevor Cai}, \bibinfo{person}{Eliza Rutherford}, \bibinfo{person}{Diego de Las~Casas}, \bibinfo{person}{Lisa~Anne Hendricks}, \bibinfo{person}{Johannes Welbl}, \bibinfo{person}{Aidan Clark}, \bibinfo{person}{Tom Hennigan}, \bibinfo{person}{Eric Noland}, \bibinfo{person}{Katie Millican}, \bibinfo{person}{George van~den Driessche}, \bibinfo{person}{Bogdan Damoc}, \bibinfo{person}{Aurelia Guy}, \bibinfo{person}{Simon Osindero}, \bibinfo{person}{Karen Simonyan}, \bibinfo{person}{Erich Elsen}, \bibinfo{person}{Oriol Vinyals}, \bibinfo{person}{Jack~W. Rae}, {and} \bibinfo{person}{Laurent Sifre}.} \bibinfo{year}{2022}\natexlab{}.
\newblock \showarticletitle{Training compute-optimal large language models}. In \bibinfo{booktitle}{\emph{Proceedings of the 36th International Conference on Neural Information Processing Systems}} \emph{(\bibinfo{series}{NIPS '22})}. Article \bibinfo{articleno}{2176}, \bibinfo{numpages}{15}~pages.
\newblock
\showISBNx{9781713871088}


\bibitem[Hofst{\"{a}}tter et~al\mbox{.}(2020a)]%
        {cross_architecture_knowledge_distillation}
\bibfield{author}{\bibinfo{person}{Sebastian Hofst{\"{a}}tter}, \bibinfo{person}{Sophia Althammer}, \bibinfo{person}{Michael Schr{\"{o}}der}, \bibinfo{person}{Mete Sertkan}, {and} \bibinfo{person}{Allan Hanbury}.} \bibinfo{year}{2020}\natexlab{a}.
\newblock \showarticletitle{Improving Efficient Neural Ranking Models with Cross-Architecture Knowledge Distillation}.
\newblock \bibinfo{journal}{\emph{CoRR}}  \bibinfo{volume}{abs/2010.02666} (\bibinfo{year}{2020}).
\newblock
\showeprint[arXiv]{2010.02666}
\urldef\tempurl%
\url{https://arxiv.org/abs/2010.02666}
\showURL{%
\tempurl}


\bibitem[Hofst{\"{a}}tter et~al\mbox{.}(2020b)]%
        {Margin_MSE_loss}
\bibfield{author}{\bibinfo{person}{Sebastian Hofst{\"{a}}tter}, \bibinfo{person}{Sophia Althammer}, \bibinfo{person}{Michael Schr{\"{o}}der}, \bibinfo{person}{Mete Sertkan}, {and} \bibinfo{person}{Allan Hanbury}.} \bibinfo{year}{2020}\natexlab{b}.
\newblock \showarticletitle{Improving Efficient Neural Ranking Models with Cross-Architecture Knowledge Distillation}.
\newblock \bibinfo{journal}{\emph{CoRR}}  \bibinfo{volume}{abs/2010.02666} (\bibinfo{year}{2020}).
\newblock
\showeprint[arXiv]{2010.02666}
\urldef\tempurl%
\url{https://arxiv.org/abs/2010.02666}
\showURL{%
\tempurl}


\bibitem[Izacard et~al\mbox{.}(2022)]%
        {contriever_unsupervised_contrastive_learning}
\bibfield{author}{\bibinfo{person}{Gautier Izacard}, \bibinfo{person}{Mathilde Caron}, \bibinfo{person}{Lucas Hosseini}, \bibinfo{person}{Sebastian Riedel}, \bibinfo{person}{Piotr Bojanowski}, \bibinfo{person}{Armand Joulin}, {and} \bibinfo{person}{Edouard Grave}.} \bibinfo{year}{2022}\natexlab{}.
\newblock \showarticletitle{Unsupervised Dense Information Retrieval with Contrastive Learning}.
\newblock \bibinfo{journal}{\emph{Transactions on Machine Learning Research}} (\bibinfo{year}{2022}).
\newblock
\urldef\tempurl%
\url{https://openreview.net/forum?id=jKN1pXi7b0}
\showURL{%
\tempurl}


\bibitem[J{\'{e}}gou et~al\mbox{.}(2011)]%
        {product_quantization_nearest_neighbor}
\bibfield{author}{\bibinfo{person}{Herv{\'{e}} J{\'{e}}gou}, \bibinfo{person}{Matthijs Douze}, {and} \bibinfo{person}{Cordelia Schmid}.} \bibinfo{year}{2011}\natexlab{}.
\newblock \showarticletitle{Product Quantization for Nearest Neighbor Search}.
\newblock \bibinfo{journal}{\emph{IEEE Transactions on Pattern Analysis and Machine Intelligence}} \bibinfo{volume}{33}, \bibinfo{number}{1} (\bibinfo{year}{2011}), \bibinfo{pages}{117--128}.
\newblock
\href{https://doi.org/10.1109/TPAMI.2010.57}{doi:\nolinkurl{10.1109/TPAMI.2010.57}}


\bibitem[Kainen and Kůrková(1993)]%
        {nearly_orthogonal}
\bibfield{author}{\bibinfo{person}{Paul~C. Kainen} {and} \bibinfo{person}{Vĕra Kůrková}.} \bibinfo{year}{1993}\natexlab{}.
\newblock \showarticletitle{Quasiorthogonal dimension of euclidean spaces}.
\newblock \bibinfo{journal}{\emph{Applied Mathematics Letters}} \bibinfo{volume}{6}, \bibinfo{number}{3} (\bibinfo{year}{1993}), \bibinfo{pages}{7--10}.
\newblock
\showISSN{0893-9659}
\href{https://doi.org/10.1016/0893-9659(93)90023-G}{doi:\nolinkurl{10.1016/0893-9659(93)90023-G}}


\bibitem[Kaplan et~al\mbox{.}(2020)]%
        {scaling_laws_neural_language_models}
\bibfield{author}{\bibinfo{person}{Jared Kaplan}, \bibinfo{person}{Sam McCandlish}, \bibinfo{person}{Tom Henighan}, \bibinfo{person}{Tom~B Brown}, \bibinfo{person}{Benjamin Chess}, \bibinfo{person}{Rewon Child}, \bibinfo{person}{Scott Gray}, \bibinfo{person}{Alec Radford}, \bibinfo{person}{Jeffrey Wu}, {and} \bibinfo{person}{Dario Amodei}.} \bibinfo{year}{2020}\natexlab{}.
\newblock \showarticletitle{Scaling laws for neural language models}.
\newblock \bibinfo{journal}{\emph{arXiv preprint arXiv:2001.08361}} (\bibinfo{year}{2020}).
\newblock


\bibitem[Karpukhin et~al\mbox{.}(2020)]%
        {dense_passage_retrieval}
\bibfield{author}{\bibinfo{person}{Vladimir Karpukhin}, \bibinfo{person}{Barlas Oguz}, \bibinfo{person}{Sewon Min}, \bibinfo{person}{Patrick Lewis}, \bibinfo{person}{Ledell Wu}, \bibinfo{person}{Sergey Edunov}, \bibinfo{person}{Danqi Chen}, {and} \bibinfo{person}{Wen-tau Yih}.} \bibinfo{year}{2020}\natexlab{}.
\newblock \showarticletitle{Dense Passage Retrieval for Open-Domain Question Answering}. In \bibinfo{booktitle}{\emph{Proceedings of the 2020 Conference on Empirical Methods in Natural Language Processing (EMNLP)}}. \bibinfo{publisher}{Association for Computational Linguistics}, \bibinfo{pages}{6769--6781}.
\newblock
\href{https://doi.org/10.18653/v1/2020.emnlp-main.550}{doi:\nolinkurl{10.18653/v1/2020.emnlp-main.550}}


\bibitem[Killingback and Zamani(2025)]%
        {CRUMB}
\bibfield{author}{\bibinfo{person}{Julian Killingback} {and} \bibinfo{person}{Hamed Zamani}.} \bibinfo{year}{2025}\natexlab{}.
\newblock \bibinfo{title}{Benchmarking Information Retrieval Models on Complex Retrieval Tasks}.
\newblock
\showeprint[arxiv]{2509.07253}~[cs.IR]
\urldef\tempurl%
\url{https://arxiv.org/abs/2509.07253}
\showURL{%
\tempurl}


\bibitem[Killingback et~al\mbox{.}(2025)]%
        {hypencoder_hypernetworks_retrieval}
\bibfield{author}{\bibinfo{person}{Julian Killingback}, \bibinfo{person}{Hansi Zeng}, {and} \bibinfo{person}{Hamed Zamani}.} \bibinfo{year}{2025}\natexlab{}.
\newblock \showarticletitle{Hypencoder: Hypernetworks for Information Retrieval}. In \bibinfo{booktitle}{\emph{Proceedings of the 48th International ACM SIGIR Conference on Research and Development in Information Retrieval}} \emph{(\bibinfo{series}{SIGIR '25})}. \bibinfo{publisher}{Association for Computing Machinery}, \bibinfo{address}{New York, NY, USA}, \bibinfo{pages}{2372--2383}.
\newblock
\showISBNx{9798400715921}
\href{https://doi.org/10.1145/3726302.3729983}{doi:\nolinkurl{10.1145/3726302.3729983}}


\bibitem[Kusupati et~al\mbox{.}(2022)]%
        {matryoshka_representations_adaptive}
\bibfield{author}{\bibinfo{person}{Aditya Kusupati}, \bibinfo{person}{Gantavya Bhatt}, \bibinfo{person}{Aniket Rege}, \bibinfo{person}{Matthew Wallingford}, \bibinfo{person}{Aditya Sinha}, \bibinfo{person}{Vivek Ramanujan}, \bibinfo{person}{William Howard-Snyder}, \bibinfo{person}{Kaifeng Chen}, \bibinfo{person}{Sham Kakade}, \bibinfo{person}{Prateek Jain}, {et~al\mbox{.}}} \bibinfo{year}{2022}\natexlab{}.
\newblock \showarticletitle{Matryoshka Representation Learning.}. In \bibinfo{booktitle}{\emph{Advances in Neural Information Processing Systems}}.
\newblock


\bibitem[Lassance et~al\mbox{.}(2024)]%
        {splade_v3}
\bibfield{author}{\bibinfo{person}{Carlos Lassance}, \bibinfo{person}{Herv{\'{e}} D{\'{e}}jean}, \bibinfo{person}{Thibault Formal}, {and} \bibinfo{person}{St{\'{e}}phane Clinchant}.} \bibinfo{year}{2024}\natexlab{}.
\newblock \showarticletitle{SPLADE-v3: New baselines for {SPLADE}}.
\newblock \bibinfo{journal}{\emph{CoRR}}  \bibinfo{volume}{abs/2403.06789} (\bibinfo{year}{2024}).
\newblock
\showeprint[arXiv]{2403.06789}
\href{https://doi.org/10.48550/ARXIV.2403.06789}{doi:\nolinkurl{10.48550/ARXIV.2403.06789}}


\bibitem[Luan et~al\mbox{.}(2021)]%
        {sparse_dense_attentional_representations}
\bibfield{author}{\bibinfo{person}{Yi Luan}, \bibinfo{person}{Jacob Eisenstein}, \bibinfo{person}{Kristina Toutanova}, {and} \bibinfo{person}{Michael Collins}.} \bibinfo{year}{2021}\natexlab{}.
\newblock \showarticletitle{Sparse, Dense, and Attentional Representations for Text Retrieval}.
\newblock \bibinfo{journal}{\emph{Transactions of the Association for Computational Linguistics}}  \bibinfo{volume}{9} (\bibinfo{year}{2021}), \bibinfo{pages}{329--345}.
\newblock
\href{https://doi.org/10.1162/TACL_A_00369}{doi:\nolinkurl{10.1162/TACL_A_00369}}


\bibitem[Menon et~al\mbox{.}(2022)]%
        {defense_dual_encoders_ranking}
\bibfield{author}{\bibinfo{person}{Aditya Menon}, \bibinfo{person}{Sadeep Jayasumana}, \bibinfo{person}{Ankit~Singh Rawat}, \bibinfo{person}{Seungyeon Kim}, \bibinfo{person}{Sashank Reddi}, {and} \bibinfo{person}{Sanjiv Kumar}.} \bibinfo{year}{2022}\natexlab{}.
\newblock \showarticletitle{In defense of dual-encoders for neural ranking}. In \bibinfo{booktitle}{\emph{Proceedings of the 39th International Conference on Machine Learning}}. \bibinfo{publisher}{PMLR}.
\newblock
\urldef\tempurl%
\url{https://proceedings.mlr.press/v162/menon22a.html}
\showURL{%
\tempurl}


\bibitem[Nguyen et~al\mbox{.}(2016)]%
        {ms_marco_dataset}
\bibfield{author}{\bibinfo{person}{Tri Nguyen}, \bibinfo{person}{Mir Rosenberg}, \bibinfo{person}{Xia Song}, \bibinfo{person}{Jianfeng Gao}, \bibinfo{person}{Saurabh Tiwary}, \bibinfo{person}{Rangan Majumder}, {and} \bibinfo{person}{Li Deng}.} \bibinfo{year}{2016}\natexlab{}.
\newblock \showarticletitle{{MS MARCO}: A Human Generated MAchine Reading COmprehension Dataset}. In \bibinfo{booktitle}{\emph{Proceedings of the Workshop on Cognitive Computation: Integrating neural and symbolic approaches}}, Vol.~\bibinfo{volume}{1773}. \bibinfo{publisher}{CEUR-WS.org}.
\newblock
\urldef\tempurl%
\url{https://ceur-ws.org/Vol-1773/CoCoNIPS_2016_paper9.pdf}
\showURL{%
\tempurl}


\bibitem[Paszke et~al\mbox{.}(2019)]%
        {PyTorch}
\bibfield{author}{\bibinfo{person}{Adam Paszke}, \bibinfo{person}{Sam Gross}, \bibinfo{person}{Francisco Massa}, \bibinfo{person}{Adam Lerer}, \bibinfo{person}{James Bradbury}, \bibinfo{person}{Gregory Chanan}, \bibinfo{person}{Trevor Killeen}, \bibinfo{person}{Zeming Lin}, \bibinfo{person}{Natalia Gimelshein}, \bibinfo{person}{Luca Antiga}, \bibinfo{person}{Alban Desmaison}, \bibinfo{person}{Andreas K{\"{o}}pf}, \bibinfo{person}{Edward~Z. Yang}, \bibinfo{person}{Zachary DeVito}, \bibinfo{person}{Martin Raison}, \bibinfo{person}{Alykhan Tejani}, \bibinfo{person}{Sasank Chilamkurthy}, \bibinfo{person}{Benoit Steiner}, \bibinfo{person}{Lu Fang}, \bibinfo{person}{Junjie Bai}, {and} \bibinfo{person}{Soumith Chintala}.} \bibinfo{year}{2019}\natexlab{}.
\newblock \showarticletitle{PyTorch: An Imperative Style, High-Performance Deep Learning Library}. In \bibinfo{booktitle}{\emph{Advances in Neural Information Processing Systems 32: Annual Conference on Neural Information Processing Systems 2019, NeurIPS 2019, December 8-14, 2019, Vancouver, BC, Canada}}, \bibfield{editor}{\bibinfo{person}{Hanna~M. Wallach}, \bibinfo{person}{Hugo Larochelle}, \bibinfo{person}{Alina Beygelzimer}, \bibinfo{person}{Florence d'Alch{\'{e}}{-}Buc}, \bibinfo{person}{Emily~B. Fox}, {and} \bibinfo{person}{Roman Garnett}} (Eds.). \bibinfo{pages}{8024--8035}.
\newblock
\urldef\tempurl%
\url{https://proceedings.neurips.cc/paper/2019/hash/bdbca288fee7f92f2bfa9f7012727740-Abstract.html}
\showURL{%
\tempurl}


\bibitem[Portes et~al\mbox{.}(2025)]%
        {retrieval_capabilities_scaling_flops}
\bibfield{author}{\bibinfo{person}{Jacob Portes}, \bibinfo{person}{Connor Jennings}, \bibinfo{person}{Erica Ji}, \bibinfo{person}{Yuen Sasha~Doubov}, {and} \bibinfo{person}{Michael Carbin}.} \bibinfo{year}{2025}\natexlab{}.
\newblock \bibinfo{title}{Retrieval Capabilities of Large Language Models Scale with Pretraining FLOPs}.
\newblock
\showeprint[arxiv]{2508.17400}~[cs.LG]
\urldef\tempurl%
\url{https://arxiv.org/abs/2508.17400}
\showURL{%
\tempurl}


\bibitem[Prakash et~al\mbox{.}(2021)]%
        {rance}
\bibfield{author}{\bibinfo{person}{Prafull Prakash}, \bibinfo{person}{Julian Killingback}, {and} \bibinfo{person}{Hamed Zamani}.} \bibinfo{year}{2021}\natexlab{}.
\newblock \showarticletitle{Learning Robust Dense Retrieval Models from Incomplete Relevance Labels}. In \bibinfo{booktitle}{\emph{{SIGIR} '21: The 44th International {ACM} {SIGIR} Conference on Research and Development in Information Retrieval, Virtual Event, Canada, July 11-15, 2021}}, \bibfield{editor}{\bibinfo{person}{Fernando Diaz}, \bibinfo{person}{Chirag Shah}, \bibinfo{person}{Torsten Suel}, \bibinfo{person}{Pablo Castells}, \bibinfo{person}{Rosie Jones}, {and} \bibinfo{person}{Tetsuya Sakai}} (Eds.). \bibinfo{publisher}{{ACM}}, \bibinfo{pages}{1728--1732}.
\newblock
\href{https://doi.org/10.1145/3404835.3463106}{doi:\nolinkurl{10.1145/3404835.3463106}}


\bibitem[Ren et~al\mbox{.}(2021)]%
        {rocketqav2_joint_training}
\bibfield{author}{\bibinfo{person}{Ruiyang Ren}, \bibinfo{person}{Yingqi Qu}, \bibinfo{person}{Jing Liu}, \bibinfo{person}{Wayne~Xin Zhao}, \bibinfo{person}{QiaoQiao She}, \bibinfo{person}{Hua Wu}, \bibinfo{person}{Haifeng Wang}, {and} \bibinfo{person}{Ji-Rong Wen}.} \bibinfo{year}{2021}\natexlab{}.
\newblock \showarticletitle{{R}ocket{QA}v2: A Joint Training Method for Dense Passage Retrieval and Passage Re-ranking}. In \bibinfo{booktitle}{\emph{Proceedings of the 2021 Conference on Empirical Methods in Natural Language Processing}}. \bibinfo{publisher}{Association for Computational Linguistics}, \bibinfo{pages}{2825--2835}.
\newblock
\href{https://doi.org/10.18653/v1/2021.emnlp-main.224}{doi:\nolinkurl{10.18653/v1/2021.emnlp-main.224}}


\bibitem[Santhanam et~al\mbox{.}(2022)]%
        {ColBERTv2}
\bibfield{author}{\bibinfo{person}{Keshav Santhanam}, \bibinfo{person}{Omar Khattab}, \bibinfo{person}{Jon Saad{-}Falcon}, \bibinfo{person}{Christopher Potts}, {and} \bibinfo{person}{Matei Zaharia}.} \bibinfo{year}{2022}\natexlab{}.
\newblock \showarticletitle{ColBERTv2: Effective and Efficient Retrieval via Lightweight Late Interaction}. In \bibinfo{booktitle}{\emph{Proceedings of the 2022 Conference of the North American Chapter of the Association for Computational Linguistics: Human Language Technologies, {NAACL} 2022, Seattle, WA, United States, July 10-15, 2022}}, \bibfield{editor}{\bibinfo{person}{Marine Carpuat}, \bibinfo{person}{Marie{-}Catherine de~Marneffe}, {and} \bibinfo{person}{Iv{\'{a}}n Vladimir~Meza Ru{\'{\i}}z}} (Eds.). \bibinfo{publisher}{Association for Computational Linguistics}, \bibinfo{pages}{3715--3734}.
\newblock
\href{https://doi.org/10.18653/V1/2022.NAACL-MAIN.272}{doi:\nolinkurl{10.18653/V1/2022.NAACL-MAIN.272}}


\bibitem[Vaswani et~al\mbox{.}(2017)]%
        {attention_is_all_you_need}
\bibfield{author}{\bibinfo{person}{Ashish Vaswani}, \bibinfo{person}{Noam Shazeer}, \bibinfo{person}{Niki Parmar}, \bibinfo{person}{Jakob Uszkoreit}, \bibinfo{person}{Llion Jones}, \bibinfo{person}{Aidan~N. Gomez}, \bibinfo{person}{Lukasz Kaiser}, {and} \bibinfo{person}{Illia Polosukhin}.} \bibinfo{year}{2017}\natexlab{}.
\newblock \showarticletitle{Attention is All you Need}. In \bibinfo{booktitle}{\emph{Advances in Neural Information Processing Systems 30: Annual Conference on Neural Information Processing Systems 2017, December 4-9, 2017, Long Beach, CA, {USA}}}, \bibfield{editor}{\bibinfo{person}{Isabelle Guyon}, \bibinfo{person}{Ulrike von Luxburg}, \bibinfo{person}{Samy Bengio}, \bibinfo{person}{Hanna~M. Wallach}, \bibinfo{person}{Rob Fergus}, \bibinfo{person}{S.~V.~N. Vishwanathan}, {and} \bibinfo{person}{Roman Garnett}} (Eds.). \bibinfo{pages}{5998--6008}.
\newblock
\urldef\tempurl%
\url{https://proceedings.neurips.cc/paper/2017/hash/3f5ee243547dee91fbd053c1c4a845aa-Abstract.html}
\showURL{%
\tempurl}


\bibitem[Wang et~al\mbox{.}(2023)]%
        {doris_mae}
\bibfield{author}{\bibinfo{person}{Jianyou Wang}, \bibinfo{person}{Kaicheng Wang}, \bibinfo{person}{Xiaoyue Wang}, \bibinfo{person}{Prudhviraj Naidu}, \bibinfo{person}{Leon Bergen}, {and} \bibinfo{person}{Ramamohan Paturi}.} \bibinfo{year}{2023}\natexlab{}.
\newblock \showarticletitle{Scientific Document Retrieval using Multi-level Aspect-based Queries}. In \bibinfo{booktitle}{\emph{Advances in Neural Information Processing Systems 36: Annual Conference on Neural Information Processing Systems 2023, NeurIPS 2023, New Orleans, LA, USA, December 10 - 16, 2023}}, \bibfield{editor}{\bibinfo{person}{Alice Oh}, \bibinfo{person}{Tristan Naumann}, \bibinfo{person}{Amir Globerson}, \bibinfo{person}{Kate Saenko}, \bibinfo{person}{Moritz Hardt}, {and} \bibinfo{person}{Sergey Levine}} (Eds.).
\newblock
\urldef\tempurl%
\url{http://papers.nips.cc/paper\_files/paper/2023/hash/78f9c04bdcb06f1ada3902912d8b64ba-Abstract-Datasets\_and\_Benchmarks.html}
\showURL{%
\tempurl}


\bibitem[Wang et~al\mbox{.}(2021)]%
        {anns_comprehensive_survey}
\bibfield{author}{\bibinfo{person}{Mengzhao Wang}, \bibinfo{person}{Xiaoliang Xu}, \bibinfo{person}{Qiang Yue}, {and} \bibinfo{person}{Yuxiang Wang}.} \bibinfo{year}{2021}\natexlab{}.
\newblock \showarticletitle{A comprehensive survey and experimental comparison of graph-based approximate nearest neighbor search}.
\newblock \bibinfo{journal}{\emph{Proc. VLDB Endow.}} \bibinfo{volume}{14}, \bibinfo{number}{11} (\bibinfo{date}{July} \bibinfo{year}{2021}), \bibinfo{pages}{1964–1978}.
\newblock
\showISSN{2150-8097}
\href{https://doi.org/10.14778/3476249.3476255}{doi:\nolinkurl{10.14778/3476249.3476255}}


\bibitem[Weller et~al\mbox{.}(2025a)]%
        {theoretical_limitations_embedding_retrieval}
\bibfield{author}{\bibinfo{person}{Orion Weller}, \bibinfo{person}{Michael Boratko}, \bibinfo{person}{Iftekhar Naim}, {and} \bibinfo{person}{Jinhyuk Lee}.} \bibinfo{year}{2025}\natexlab{a}.
\newblock \showarticletitle{On the Theoretical Limitations of Embedding-Based Retrieval}.
\newblock \bibinfo{journal}{\emph{CoRR}}  \bibinfo{volume}{abs/2508.21038} (\bibinfo{year}{2025}).
\newblock
\showeprint[arXiv]{2508.21038}
\href{https://doi.org/10.48550/arXiv.2508.21038}{doi:\nolinkurl{10.48550/arXiv.2508.21038}}


\bibitem[Weller et~al\mbox{.}(2025b)]%
        {Ettin_suite}
\bibfield{author}{\bibinfo{person}{Orion Weller}, \bibinfo{person}{Kathryn Ricci}, \bibinfo{person}{Marc Marone}, \bibinfo{person}{Antoine Chaffin}, \bibinfo{person}{Dawn~J. Lawrie}, {and} \bibinfo{person}{Benjamin~Van Durme}.} \bibinfo{year}{2025}\natexlab{b}.
\newblock \showarticletitle{Seq vs Seq: An Open Suite of Paired Encoders and Decoders}.
\newblock \bibinfo{journal}{\emph{CoRR}}  \bibinfo{volume}{abs/2507.11412} (\bibinfo{year}{2025}).
\newblock
\showeprint[arXiv]{2507.11412}
\href{https://doi.org/10.48550/ARXIV.2507.11412}{doi:\nolinkurl{10.48550/ARXIV.2507.11412}}


\bibitem[Wolf et~al\mbox{.}(2020)]%
        {HuggingFace}
\bibfield{author}{\bibinfo{person}{Thomas Wolf}, \bibinfo{person}{Lysandre Debut}, \bibinfo{person}{Victor Sanh}, \bibinfo{person}{Julien Chaumond}, \bibinfo{person}{Clement Delangue}, \bibinfo{person}{Anthony Moi}, \bibinfo{person}{Pierric Cistac}, \bibinfo{person}{Tim Rault}, \bibinfo{person}{R{\'{e}}mi Louf}, \bibinfo{person}{Morgan Funtowicz}, {and} \bibinfo{person}{Jamie Brew}.} \bibinfo{year}{2020}\natexlab{}.
\newblock \showarticletitle{HuggingFace's Transformers: State-of-the-art Natural Language Processing}. In \bibinfo{booktitle}{\emph{Proceedings of the 2020 Conference on Empirical Methods in Natural Language Processing: System Demonstrations}}. \bibinfo{publisher}{Association for Computational Linguistics}, \bibinfo{address}{Online}, \bibinfo{pages}{38--45}.
\newblock
\href{https://doi.org/10.18653/v1/2020.emnlp-demos.6}{doi:\nolinkurl{10.18653/v1/2020.emnlp-demos.6}}


\bibitem[Xiong et~al\mbox{.}(2021)]%
        {ance_negative_contrastive_learning}
\bibfield{author}{\bibinfo{person}{Lee Xiong}, \bibinfo{person}{Chenyan Xiong}, \bibinfo{person}{Ye Li}, \bibinfo{person}{Kwok-Fung Tang}, \bibinfo{person}{Jialin Liu}, \bibinfo{person}{Paul Bennett}, \bibinfo{person}{Junaid Ahmed}, {and} \bibinfo{person}{Arnold Overwijk}.} \bibinfo{year}{2021}\natexlab{}.
\newblock \showarticletitle{Approximate Nearest Neighbor Negative Contrastive Learning for Dense Text Retrieval}. In \bibinfo{booktitle}{\emph{Proceedings of the International Conference on Learning Representations (ICLR)}}.
\newblock


\bibitem[Yang et~al\mbox{.}(2020)]%
        {redundancy_elimination_dense_vectors}
\bibfield{author}{\bibinfo{person}{Xueguang Yang}, \bibinfo{person}{Ronak Zhong}, \bibinfo{person}{Tianqi Fang}, {and} \bibinfo{person}{Jimmy Lin}.} \bibinfo{year}{2020}\natexlab{}.
\newblock \showarticletitle{Simple and Effective Unsupervised Redundancy Elimination to Compress Dense Vectors for Passage Retrieval}. In \bibinfo{booktitle}{\emph{Proceedings of the 2020 Conference on Empirical Methods in Natural Language Processing (EMNLP)}}. \bibinfo{publisher}{Association for Computational Linguistics}, \bibinfo{pages}{3154--3159}.
\newblock
\urldef\tempurl%
\url{https://aclanthology.org/2020.emnlp-main.255}
\showURL{%
\tempurl}


\bibitem[Zamani et~al\mbox{.}(2018)]%
        {SNRM}
\bibfield{author}{\bibinfo{person}{Hamed Zamani}, \bibinfo{person}{Mostafa Dehghani}, \bibinfo{person}{W.~Bruce Croft}, \bibinfo{person}{Erik Learned-Miller}, {and} \bibinfo{person}{Jaap Kamps}.} \bibinfo{year}{2018}\natexlab{}.
\newblock \showarticletitle{From Neural Re-Ranking to Neural Ranking: Learning a Sparse Representation for Inverted Indexing}. In \bibinfo{booktitle}{\emph{Proceedings of the 27th ACM International Conference on Information and Knowledge Management}} (Torino, Italy) \emph{(\bibinfo{series}{CIKM '18})}. \bibinfo{publisher}{Association for Computing Machinery}, \bibinfo{address}{New York, NY, USA}, \bibinfo{pages}{497–506}.
\newblock
\showISBNx{9781450360142}
\href{https://doi.org/10.1145/3269206.3271800}{doi:\nolinkurl{10.1145/3269206.3271800}}


\bibitem[Zeng et~al\mbox{.}(2025)]%
        {scaling_sparse_dense_decoder_llms}
\bibfield{author}{\bibinfo{person}{Hansi Zeng}, \bibinfo{person}{Julian Killingback}, {and} \bibinfo{person}{Hamed Zamani}.} \bibinfo{year}{2025}\natexlab{}.
\newblock \showarticletitle{Scaling Sparse and Dense Retrieval in Decoder-Only LLMs}. In \bibinfo{booktitle}{\emph{Proceedings of the 48th International ACM SIGIR Conference on Research and Development in Information Retrieval}} (Padua, Italy) \emph{(\bibinfo{series}{SIGIR '25})}. \bibinfo{publisher}{Association for Computing Machinery}, \bibinfo{address}{New York, NY, USA}, \bibinfo{pages}{2679–2684}.
\newblock
\showISBNx{9798400715921}
\href{https://doi.org/10.1145/3726302.3730225}{doi:\nolinkurl{10.1145/3726302.3730225}}


\bibitem[Zheng et~al\mbox{.}(2025)]%
        {legal_retrieval_benchmark}
\bibfield{author}{\bibinfo{person}{Lucia Zheng}, \bibinfo{person}{Neel Guha}, \bibinfo{person}{Javokhir Arifov}, \bibinfo{person}{Sarah Zhang}, \bibinfo{person}{Michal Skreta}, \bibinfo{person}{Christopher~D. Manning}, \bibinfo{person}{Peter Henderson}, {and} \bibinfo{person}{Daniel~E. Ho}.} \bibinfo{year}{2025}\natexlab{}.
\newblock \showarticletitle{A Reasoning-Focused Legal Retrieval Benchmark}. In \bibinfo{booktitle}{\emph{Proceedings of the 2025 Symposium on Computer Science and Law, {CSLAW} 2025, Munich, Germany, March 25-27, 2025}}. \bibinfo{publisher}{{ACM}}, \bibinfo{pages}{169--193}.
\newblock
\href{https://doi.org/10.1145/3709025.3712219}{doi:\nolinkurl{10.1145/3709025.3712219}}


\bibitem[Zipf(1949)]%
        {zipfian_distribution}
\bibfield{author}{\bibinfo{person}{George~K. Zipf}.} \bibinfo{year}{1949}\natexlab{}.
\newblock \bibinfo{booktitle}{\emph{Human Behaviour and the Principle of Least Effort}}.
\newblock \bibinfo{publisher}{Addison-Wesley}.
\newblock


\end{thebibliography}

\appendix

\end{document}